\documentclass[showpacs,amsmath,amssymb,nofootinbib,prd]{revtex4}

\usepackage{graphicx}
\usepackage{enumerate}

\newcommand{\beq}{\begin{equation}}
\newcommand{\eeq}{\end{equation}}
\newcommand{\bea}{\begin{eqnarray}}
\newcommand{\eea}{\end{eqnarray}}

\begin{document}

\title{The general Two-Higgs doublet eXtensions of the SM: a saucerful of
secrets}
\author{J.L. D\'{\i}az-Cruz}
\email{ldiaz@sirio.ifuap.buap.mx} \affiliation{Cuerpo Acad\'emico de
Part\'{\i}culas, Campos y Relatividad Facultad de Ciencias
F\'{\i}sico-Matem\'aticas, BUAP. Apdo. Postal 1364, C.P. 72000
Puebla, Pue., M\'exico}
\author{A. D\'iaz-Furlong}
\email{adiazfurlong@yahoo.com} \affiliation{Cuerpo Acad\'emico de
Part\'{\i}culas, Campos y Relatividad Facultad de Ciencias
F\'{\i}sico-Matem\'aticas, BUAP. Apdo. Postal 1364, C.P. 72000
Puebla, Pue., M\'exico}
\author{J.H. Montes de Oca Y.}
\email{halim@esfm.ipn.mx} \affiliation{Cuerpo Acad\'emico de
Part\'{\i}culas, Campos y Relatividad Facultad de Ciencias
F\'{\i}sico-Matem\'aticas, BUAP. Apdo. Postal 1364, C.P. 72000
Puebla, Pue., M\'exico}

\begin{abstract}
We discuss the most general formulation of the Two-Higgs doublet
model, which incorporates flavor changing neutral scalar
interactions (FCNSI) and CP violation (CPV) from several sources. CP
violation can arise either from Yukawa terms or from the Higgs
potential, be it explicit or spontaneous. We show how the model,
which is denoted as 2HDM-X, reduces to some versions known in the
literature (2HDM-I,II,III), as well as some of their variants (top,
lepton, dark) denoted here as 2HDM-IV. We also discuss another limit
that includes CPV and Yukawa four textures to control FCNSI, which
we denote as 2HDM-V.
We evaluate the CPV asymmetry for the decay $h\to bcW$, which may
allow to test the patterns of FCNSI and CPV, that arises in these
models.
\end{abstract}

 \maketitle

\section{Introduction}

Despite the success of the Standard Model (SM) in the gauge and
fermion sectors, the Higgs sector remains the least tested aspect of
the model \cite{Gunion:1989we}, which leaves the puzzles associated
with the mechanism of electroweak symmetry breaking (EWSB) still
unsolved. On one hand, the analysis of raditive corrections within
the SM
\cite{Erler:2010wa,Flacher:2008zq,DiazCruz:2003qs,Baur:2002gp},
points towards the existence of a Higgs boson, with a mass of the
order of the EW scale, which in turn could be detected at the LHC
\cite{Carena:2002es,Nath:2010zj}. On the other hand, the SM is often
considered as an effective theory, valid up to an energy scale of
$O(TeV)$, that eventually will be replaced by a more fundamental
theory \cite{Bustamante:2009us}, which will explain, among other
things, the physics behind EWSB and perhaps even the origin of
flavor. Many examples of candidate theories, which range from
supersymmetry \cite{Ellis:2010wx,susyrev,Haber:2000jh} to strongly
interacting models \cite{ArkaniHamed:2001nc,Aranda:2007tg} as well
as some extra dimensional scenarios
\cite{Aranda:2002dz,Chang:2010et,Iltan:2007zz}, include a
multi-scalar Higgs sector. In particular, models with two scalar
doublets have been studied extensively
\cite{Haber:1978jt,Liu:1987ng,Wu:1994ja}, as they include a rich
structure with interesting phenomenology
\cite{Carena:2000yx,Ginzburg:2002wt,Ginzburg:2004vp}.

Several versions of the 2HDM have been studied in the literature
\cite{Accomando:2006ga}. Some models (known as 2HDM-I and 2HDM-II)
involve natural flavor conservation \cite{Glashow:1976nt}, while
other models (known as 2HDM-III)  \cite{Accomando:2006ga}, allow for
the presence of flavor changing scalar interactions (FCNSI) at a
level consistent with low-energy constraints \cite{DiazCruz:2004tr}.
There are also some variants (known as top, lepton, neutrino), where
one Higgs doublet couples predominantly to one type of fermion
\cite{Atwood:2005bf}, while in other models it is even possible to
identify a candidate for dark matter \cite{DiazCruz:2007be}.
 The definition of all these models, depends on the Yukawa structure
and symmetries of the Higgs sector
\cite{DiazCruz:2004ss,Aranda:2005st,DiazCruz:2006ki,Barbieri:2005kf,Froggatt:2007qp},
whose origin is still not known. The possible appearance of new
sources of CP violation is another characteristic of these models
\cite{Ginzburg:2005yw}.

In this paper we aim to discuss the most general version of the
Two-Higgs doublet model (2HDM), which incorporates flavor or CP
violation from all possible sources
\cite{Maniatis:2007vn,Gerard:2007kn,ElKaffas:2007rq}. We also
discuss how the general model, denoted here as 2HDM-X, reduces in
certain limits to the versions known as 2HDM-I,II,III, as well as
some of their variants which we shall name as 2HDM-IV and 2HDM-V.
The logic of the naming scheme that we adopt here, consists in
identifying distinctive physical characteristics that can be
associated with the models and have sufficient merit to single out
them.

Within model I (2HDM-I) where only one Higgs doublet generates all
gauge and fermion masses \cite{Carena:2002es}, while the second
doublet only knows about this through mixing, and thus the Higgs
phenomenology will share some similarities with the SM, although the
SM Higgs couplings will now be shared among the neutral scalar
spectrum. The presence of a charged Higgs boson is clearly the
signal beyond the SM. Within 2DHM-II one also has natural flavor
conservation \cite{Glashow:1976nt}, and its phenomenology will be
similar to the 2HDM-I, although in this case the SM couplings are
shared not only because of mixing, but also because of the Yukawa
structure. On the other hand, the distinctive characteristic of
2HDM-III is the presence of FCNSI, which require a certain mechanism
in order to suppress them, for instance one can imposes a certain
texture for the Yukawa couplings \cite{Fritzsch:1977za}, which will
then predict a pattern of FCNSI Higgs couplings \cite{Cheng:1987rs}.
Within all those models (2HDM I,II,III)
\cite{Carcamo:2006dp,Zhou:2003kd,Aoki:2009ha}, the Higgs doublets
couple, in principle, with all fermion families, with a strength
proportional to the fermion masses, modulo other parameters.

There are also other models where the Higgs doublets couple
non-universally to the fermion families, which have also been
discussed in the literature
\cite{Atwood:2005bf,Logan:2010ag,Logan:2009uf}, we shall denote this
class of family non-universal models as 2HDM-IV. In principle, the
general model includes CPV, which could arise from the same CPV
phase that appears in the CKM matrix, as in the SM, from some other
extra phase coming from the Yukawa sector or from the Higgs
potential \cite{Gunion:2005ja}. However, in order to discuss which
type of CP violation can appear in each case, we shall use the label
2HDM-V to denote the class of models which, besides containing a
generic pattern of FCNSI, moduled by certain texture, will include
new sources of CPV as well.

Our formulation of the 2HDM-X, which is discussed in detail in
section 2, relies on the Higgs mass-eigenstates basis
\cite{DiazCruz:1992uw}. It seems to us that this is more appropriate
in order to relate the low-energy constraints on the parameters of
the models, with the predicted high energy signatures to be searched
at future colliders.
The different models can be characterized by a set of invariants,
signaling the possible appearence of CP violation
\cite{Ginzburg:2005yw}, either from the bosonic or fermionic
sectors. Section 3 contains the discussion of general Yukawa
couplings, with spontaneous or explicit CPV parameters for the Higgs
bosons. We then show in section 4, our evaluation of the CPV
asymmetry for the decay $h\to bcW$, which may allow to test these
patterns of FCNSI and CPV at a future colliders. Finally our
conclusions are included in section 5, while some technical details
are included in the appendices.

\section{A General formulation of the 2HDM and its limiting cases.}

The Two-Higgs doublet extension of the SM includes two scalar
doublets of equal hypercharge, denoted by:
$\Phi_{1,2}=(\phi^+_{1,2}, \phi^0_{1,2})^T$. Depending on the Yukawa
matrices $Y^q_{1,2}$ ($q=u,$ $d$) that are allowed, one defines the
particular versions of the 2HDM. FCNSI appear at tree level when
more than one Higgs doublet couples to both types of quarks ($u$ and
$d$) \cite{susyrev}, and a certain mechanism should be invoked in
order to bring them under control. The CP properties of the Higgs
boson depend on the symmetries of the potential
\cite{Gunion:2005ja}.

In order to clarify the discussion of the many models that have been
presented in the literature, we shall present in the next
subsections, a classification scheme for these models, which have
different patterns of FCNSI and CP properties. We shall discuss in
this paper first the most general formulation of the 2HDM-X, and
will consider the 2HDM versions usually discussed in the literature,
which are known as 2HDM-I,II,III, as well as some variants
(2HDM-IV). We shall discuss then in detail, some cases that have not
been considered before, which we denote as 2HDM-V. Although the
2HDM-X suffers from the FCNSI problem, we shall discuss it first in
general terms, without referring the specific mechanism that is used
to address the problem, which will be done later in this section.

\subsection{A classification of models}

We shall define here the different types of models according to
their Yukawa structure, the Hermiticity of the Yukawa matrices and
the CP properties of the bosonic Higgs sector. Thus, the most
general version of the 2HDM is defined through the following
assumptions:

\begin{enumerate}[i)]
\item In principle we allow each Higgs doublet to couple to both type of
fermions, expecting  that some particular structure of the Yukawa
matrices is responsible for the suppression of flavor changing
neutral scalar interactions (FCNSI).

\item The Yukawa matrices are allowed in general to be non-Hermitian, i. e.,
$Y_{fi}\neq Y_{fi}^\dagger$ ($f=$u, d, l, $i=1,2$). The limit when
the Yukawa matrices are hermitic defines a particular version of the
models.

\item The Higgs potential admits in principle both spontaneous or
explicit CPV.
\end{enumerate}

Then the known limiting models (2HDM I,II,III), are obtained by
relaxing some of those assumptions, namely:

\begin{enumerate}
\item The 2HDM-I \cite{Lee:1973iz,Haber:1978jt,Hall:1981bc} is defined by considering that only one Higgs doublet
generates the masses of all types of fermions, as it happens in the
SM. This type of model can be obtained by assuming an additional
$Z_2$ discrete symmetry. Under a variant of this model, where the
second doublet does not mix with the first doublet, it is possible
to identify a neutral scalar as a dark matter candidate
\cite{Accomando:2006ga}, which makes it very attractive.

\item For the so called 2HDM-II \cite{Donoghue:1978cj,Haber:1978jt,Hall:1981bc}, each Higgs doublet couples only to one type
of quark, and then FCNSI do not appear at tree level. A variant of
the $Z_2$ discrete symmetry is considered here, similarly to the
case of the 2HDM-I. Two limiting cases can be considered, namely:
the 2HDM-IIa, with CP conserving Higgs sector, and THDM-IIb where
the Higgs sector ir CP violating \cite{Osland:2008aw}. This model is
also attractive because it corresponds to the Higgs sector of the
Minimal supersymmetric standard model (MSSM), at tree-level
\cite{Carena:2002es}.

\item Within the 2HDM-III \cite{Cheng:1987rs}, one considers all possible couplings among the
Higgs doublets and fermions in the Yukawa sector; thus, it is
possible to have FCNSI in this case. According to the extended
classification that we try to motivate here, we shall also assume
that within 2HDM-III the Yukawa matrices are Hermitic, whereas the
Higgs potential is CP conserving. Thus, CP violation only arise from
the CKM phase. A particular version of this model, widely studied in
the literature, assumes Hermitic Yukawa matrices with 4-textures
\cite{Fritzsch:1977za}, which has under control the FCNSI problem
\cite{Cheng:1987rs}. It also happens that when one considers loop
effects within the MSSM, its Higgs sector also becomes of type III
\cite{DiazCruz:2002er}, which again makes attractive this version of
the 2HDM. Although it is not often explicitly stated, we shall
consider that within 2HDM-III the Higgs doublets couple in principle
with all three families of quarks and leptons.

\item Family non-universal assignments are also possible, for instance we
can have models where one doublet couples to all types of quarks,
but a second doublet only couples to the 3rd family
\cite{Atwood:2005bf}. Several possibilities have been considered in
the literature, which we denote as 2HDM-IV, depending on whether the
second doublet couples to the whole third family or only to the top
(2HDM-IV-t) \cite{Logan:2010ag,Logan:2009uf}. We shall also include
within this category, those models where one doublet couples only to
charged leptons or to neutrinos \cite{Logan:2010ag,Logan:2009uf}.
\end{enumerate}

The properties of these models are summarized in tables
\ref{table-0}, \ref{table-1}. Table \ref{table-1} shows the
assumptions for the different types of the 2HDM which are considered
in this work. Besides the cases, that have been discussed in the
literature, we can define still another class of models, which we
denote as 2HDM-V, where besides having FCNSI at tree level, also
include extra sources of CP violation, either from the Yukawa or the
Higgs sectors. Within this class, we shall consider the following
sub-cases:

\begin{enumerate}[(a)]
\item The 2HDM-Va has Hermitian Yukawa matrices, but the Higgs sector is CP violating.
To work out a concrete example we shall also consider a four-texture
for the Yukawa matrices.

\item For the 2HDM-Vb we will not assume Hermiticity in the Yukawa matrices, while the Higgs sector is
CP conserving. Again, to work out an specific example we shall
consider the four-texture case for the Yukawa matrices.

%
\end{enumerate}

\begin{center}
\begin{table}
\begin{tabular}{|c|c|c|c|c|}
\hline Model type & Up quarks  & Down quarks & Charged leptons &
Neutral leptons\\ \hline
2HDM-I & $H_1$ & $H_1$ & $H_1$ & $H_1$ \\
2HDM-II & $H_2$ & $H_1$ & $H_1$ & $H_2$ \\
2HDM-III & $H_{1,2}$ & $H_{1,2}$ & $H_{1,2}$ & $H_{1,2}$ \\
2HDM-IV & $H_1$ & $H_1$ & $H_1$ & $H_2$ \\ \hline
\end{tabular}
\caption{Higgs interaction with fermions for 2HDM types.}
\label{table-0}
\end{table}
\end{center}

\begin{center}
\begin{table}
\begin{tabular}{|c|c|c|c|}
\hline Model type & FCNC & Hermiticity & Higgs sector CP
\\ \hline
I & x & -- & -- \\
II & x & -- & -- \\
III & $\checkmark$ & $\checkmark$ & $\checkmark$ (CKM) \\
Va & $\checkmark$ & $\checkmark$  & x \\
Vb & $\checkmark$ & x & $\checkmark$  \\
\hline
\end{tabular}
\caption{Symmetries under the different types of 2HDM's}
\label{table-1}
\end{table}
\end{center}
\begin{center}
\begin{table}
\begin{tabular}{|c|c|c|c|}
\hline Model type & $1^{\textrm{st}}$ family & $2^{\textrm{nd}}$
family & $3^{\textrm{rd}}$ family\\ \hline
I & $H_1$ & $H_1$ & $H_1$ \\
\hline
II & $H_2\rightarrow u$ quarks & $H_2\rightarrow u$ quarks & $H_2\rightarrow u$ quarks \\
& $H_1\rightarrow d$ quarks and leptons & $H_1\rightarrow d$ quarks and leptons & $H_2\rightarrow u$ quarks and leptons \\
\hline III & $H_1$ and $H_2$ & $H_1$ and $H_2$ & $H_1$ and $H_2$ \\
\hline
IV-$t$, $b$, $\tau$ & $H_1$ & $H_1$ & $H_2$ \\
\hline
V & $H_1$ and  $H_2$& $H_1$ and  $H_2$ & $H_1$ and  $H_2$ \\
\hline
\end{tabular}
\caption{Higgs couplings to the fermion families.}
\label{table-xtra}
\end{table}
\end{center}

\subsection{Solutions to the FCNC problem}

When both Higgs doublets couple to up- and down-type fermions, FCNSI
are allowed \cite{susyrev}. An acceptable suppression for FCNSI can
be achieved with the following mechanisms:

\begin{itemize}
\item \textit{Universal Yukawa textures.} Suppression for FCNC
can be achieved when a certain form of the Yukawa matrices that
reproduce the observed fermion masses and mixing angles is
implemented in the model. This could be done either by implementing
the Frogart-Nielsen mechanism to generate the fermion mass
hierarchies \cite{FN}, or by studying a certain ansatz for the
fermion mass matrices \cite{Fritzsch:1977za}. The first proposal for
the Higgs boson couplings \cite{Cheng:1987rs}, the so called
Cheng-Sher ansazt, was based on the Fritzsch six-texture form of the
mass matrices, namely:
\begin{equation}
M_{l}=\left(
\begin{array}{ccc}
0 & C_{q} & 0 \\
C_{q}^{\ast } & 0 & B_{q} \\
0 & B_{q}^{\ast } & A_{q}%
\end{array}%
\right) .
\end{equation}%
Then, by assuming that each Yukawa matrix $Y_{1,2}^{q}$ has the same
hierarchy, one finds: $A_{q}\simeq m_{q_{3}}$, $B_{q}\simeq \sqrt{%
m_{q_{2}}m_{q_{3}}}$ and $C_{q}\simeq \sqrt{m_{q_{1}}m_{q_{2}}}$.
Then, the fermion-fermion$^{\prime }$-Higgs boson ($f f^{\prime 0}$)
couplings obey the following pattern: $Hf_{i}f_{j} \sim
\sqrt{m_{f_i}m_{f_j}} / m_{W}$, which is also known as the
Cheng-Sher ansatz. This brings under control the FCNC problem, and
it has been extensively studied in the literature to search for
flavor-violating signals in the Higgs sector 

In our previous work we considered in detail the case of universal
four-texture Yukawa matrices \cite{DiazCruz:2004tr}, and derived the
scalar-fermion interactions, showing that it was possible to satisfy
current constraints from LFV and FCNC \cite{ourwork1,ourwork2}.
Predictions for Higgs phenomenology at the LHC was also studied in
ref. \cite{ourwork3,otherswork1}. We can consider this  a universal
model, in the sense that it was assumed that each Yukawa matrix
$Y^q_{1,2}$ has the same hierarchy.

%
%
\item \textit{Radiative Suppression of FCNC.} One could keep FCNC
under control if there exists a hierarchy between $Y^{u,d}_1$ and $%
Y^{u,d}_2$. Namely, a given set of Yukawa matrices is present at
tree-level, but the other ones arise only as a radiative effect.
This occurs for instance in the MSSM, where the type-II 2HDM
structure is not protected by any symmetry, and is transformed into
a type-III 2HDM, through the loop effects of sfermions and gauginos.
Namely, the Yukawa couplings that are already present at tree-level
in the MSSM ($Y^d_1, Y^u_2$) receive radiative corrections, while
the terms ($Y^d_2, Y^u_1$) are induced at one-loop level.

\item \textit{Alignment of the Yukawa matrices} Another solution to the FCNC
problem that have been discussed recently assumes that the Yukawa
matrices could be aligned \cite{Pich:2009sp,Jung:2010ik}. However,
it seems that if such assumption holds at a high energy scale (much
above the EW scale), it no longer holds at a low-energy scale
\cite{Braeuninger:2010td}.
\end{itemize}

\section{The lagrangian for the 2HDM}

The most general structure of the Yukawa lagrangian for the quark
fields, can be written as follows:

\begin{equation}
\mathcal{L}_{Y}^{quarks}=\overline{q}_{L}^{0}Y_{1}^{D}\phi _{1}d_{R}^{0}+%
\overline{q}_{L}^{0}Y_{2}^{D}\phi _{2}d_{R}^{0}+\overline{q}_{L}^{0}Y_{1}^{U}%
\widetilde{\phi }_{1}u_{R}^{0}+\overline{q}_{L}^{0}Y_{2}^{U}\widetilde{\phi }%
_{2}u_{R}^{0}+h.c.,  \label{yukawa}
\end{equation}

where $Y_{1,2}^{U,D}$ are the $3\times 3$ Yukawa matrices, $q_{L}$
denotes the left handed quarks doublets and $u_{R}$, $d_{R}$
correspond to the right handed singlets. Here $\widetilde{\phi
}_{1,2}=i\sigma _{2}\phi _{1,2}^{\ast }$. The superscript zero means
that the quarks are weak eigenstates. After getting a correct SSB
\cite{PortugalMinHix,Ivanov:2006yq,Maniatis:2006fs, Ma:2010ya}, the
Higgs doublets are decomposed as follows:

\begin{equation}
\phi _{1}=\left(
\begin{array}{c}
\varphi _{1}^{+} \\
\frac{v_{1}+\varphi _{1}+i\chi _{1}}{\sqrt{2}}%
\end{array}%
\right) ,  \label{doblete1}
\end{equation}

\begin{equation}
\phi _{2}=\left(
\begin{array}{c}
\varphi _{2}^{+} \\
\frac{e^{i\xi }v_{2}+\varphi _{2}+i\chi _{2}}{\sqrt{2}}%
\end{array}%
\right) .  \label{doblete2}
\end{equation}

where the v.e.v.'s $v_{1}$ and $v_{2}$ are real and positive, while
the phase $\xi $\ introduces spontaneous CP violation.
%
%
Now, we transform the quarks to the mass eigenstate basis through
the rotations: $u_{L,R}=U_{L,R}u_{L,R}^{0}$ ,
$d_{L,R}=D_{L,R}d_{L,R}^{0}$, to obtain:

\begin{eqnarray}
\mathcal{L}_{Y}^{quarks} &=&\overline{u}_{L}U_{L}Y_{1}^{D}\varphi
_{1}^{+}D_{R}^{\dagger
}d_{R}+\overline{d}_{L}D_{L}Y_{1}^{D}\frac{\varphi
_{1}+i\chi _{1}}{\sqrt{2}}D_{R}^{\dagger }d_{R}  \nonumber \\
&&+\overline{u}_{L}U_{L}Y_{2}^{D}\varphi _{2}^{+}D_{R}^{\dagger }d_{R}+%
\overline{d}_{L}D_{L}Y_{2}^{D}\frac{\varphi _{2}+i\chi _{2}}{\sqrt{2}}%
D_{R}^{\dagger }d_{R} \nonumber\\
&&+\overline{u}_{L}U_{L}Y_{1}^{U}\frac{\varphi _{1}-i\chi _{1}}{\sqrt{2}}%
U_{R}^{\dagger }u_{R}-\overline{d}_{L}D_{L}^{\dagger
}Y_{1}^{U}\varphi
_{1}^{-}U_{R}^{\dagger }u_{R}  \nonumber \\
&&+\overline{u}_{L}U_{L}Y_{2}^{U}\frac{\varphi _{2}-i\chi _{2}}{\sqrt{2}}%
U_{R}u_{R}-\overline{d}_{L}D_{L}^{\dagger }Y_{2}^{U}\varphi
_{2}^{-}U_{R}^{\dagger }u_{R} \nonumber\\
&&+\overline{u}_{L}M^{U}u_{R}+\overline{d}_{L}M^{D}d_{R}+h.c.,
\end{eqnarray}

Then, the (diagonal) mass matrices are given as follows:

\begin{equation}
M^{U}=\frac{v_{1}}{\sqrt{2}}\widetilde{Y}_{1}^{U}+e^{-i\xi }\frac{v_{2}}{%
\sqrt{2}}\widetilde{Y}_{2}^{U}  \label{mu}
\end{equation}%
and
\begin{equation}
M^{D}=\frac{v_{1}}{\sqrt{2}}\widetilde{Y}_{1}^{D}+e^{i\xi }\frac{v_{2}}{%
\sqrt{2}}\widetilde{Y}_{2}^{D},  \label{md}
\end{equation}

where $\widetilde{Y}_{1,2}^U=U_LY_{1,2}^UU_R^\dagger$ and
$\widetilde{Y}_{1,2}^D=D_LY_{1,2}^D D_R^\dagger$. Then, one can
split the Yukawa couplings into the neutral and charged terms, both
for the up and down sector. The neutral couplings for the up sector
are given in terms of (four-components) Dirac spinors as follows:

\begin{eqnarray}
\mathcal{L}_{up}^{neutral}
&=&\overline{u}\widetilde{Y}_{1}^{U}\frac{\varphi
_{1}-i\chi _{1}}{\sqrt{2}}P_{R}u+\overline{u}\widetilde{Y}_{2}^{U}\frac{%
\varphi _{2}-i\chi _{2}}{\sqrt{2}}P_{R}u  \nonumber \\
&&+\overline{u}\widetilde{Y}_{1}^{U\dagger }\frac{\varphi _{1}+i\chi _{1}}{%
\sqrt{2}}P_{L}u+\overline{u}\widetilde{Y}_{2}^{U\dagger
}\frac{\varphi
_{2}+i\chi _{2}}{\sqrt{2}}P_{L}u  \nonumber \\
&&+\overline{u}M^{U}u,
\end{eqnarray}

where $P_{L,R}=\frac{\mathbb{I}\mp \gamma ^{5}}{2}$ are the chiral
operators. In order to arrive to the final form of the Yukawa
lagrangian we need to include the Higgs mass eigenstates. When one
allows for the possibility of having CP violation in the Higgs
potential, the CP even and CP-odd components get mixed
\cite{Haber:2006ue}. This CPV Higgs mixing is included as follows,

\begin{equation}
\left(
\begin{array}{c}
\varphi _{1} \\
\varphi _{2} \\
\chi _{1} \\
\chi_{2}%
\end{array}
\right) =R\left(
\begin{array}{c}
H_{1} \\
H_{2} \\
H_{3} \\
H_{4}%
\end{array}%
\right) .  \label{higgs3}
\end{equation}

with $H_{4}$ is a Goldstone boson. The matrix $R$ can be obtain when
by relating equations (\ref{doblete1}) and (\ref{doblete2}) with the
physical Higgs mass eigenstate
\begin{equation}
\Phi _{a}=\left(
\begin{array}{c}
G^{+}v_{a}+H^{+}w_{a} \\
\frac{v}{\sqrt{2}}v_{a}+\frac{1}{\sqrt{2}}\sum_{r=1}^4 \left(
q_{r1}v_{a}+q_{r2}e^{-i\theta _{23}}w_{a}\right) H_{r}%
\end{array}%
\right) ,  \label{phys-eigen}
\end{equation}

where $a=1,2$, $r=1,...,4$ and $\hat{v}_{a}$, $\hat{w}_{a}$ are the
components of the orthogonal eigenvectors of unit norm\footnote{Here
we shall follow closely the notation of Haber and O'Neil
\cite{Haber:2006ue}.}

\begin{equation}
\widehat{v}=\left(
\begin{array}{cc}
\hat{v}_{1}, & \hat{v}_{2}%
\end{array}%
\right) =\left(
\begin{array}{cc}
\cos \beta , & e^{i\xi }\sin \beta
\end{array}%
\right)   \label{v}
\end{equation}%
and%
\begin{equation}
\widehat{w}=\left(
\begin{array}{cc}
\hat{w}_{1}, & \hat{w}_{2}%
\end{array}%
\right) =\left(
\begin{array}{cc}
-e^{-i\xi }\sin \beta , & \cos \beta
\end{array}%
\right) .  \label{w}
\end{equation}%
The values of $q_{ra}$ are written as combination of the $\theta
_{ij}$, which are the mixing angles appearing in the rotation matrix
that diagonalize the mass matrix for neutral Higgs; table
\ref{table-2} shows the different values for the $q_r$'s.

\begin{center}
\begin{table}
\begin{tabular}{|c|c|c|}
\hline $r$ & $q_{r1}$ & $q_{r2}$ \\ \hline
1 & $\cos \theta_{12}\cos \theta _{13}$ & $-\sin \theta _{12}-i\cos \theta {12}\sin\theta_{13}$\\
2 & $\sin \theta _{12}\cos \theta _{13}$ & $\cos \theta _{12}-i\sin \theta {12}\sin\theta_{13}$ \\
3 & $\sin \theta _{13}$ & $i\cos \theta _{13}$ \\
4 & $i$ & 0 \\ \hline
\end{tabular}
\caption{Mixing angles for Higgs bosons which consider spontaneous
and explicit CPV \cite{Haber:2006ue}.} \label{table-2}
\end{table}
\end{center}

It is convenient to write the following relation, for $a=1,2$
\begin{equation}
\varphi _{1}+i\chi _{1}=\sum_r \left( q_{r1}\cos \beta
-q_{r2}e^{-i\left( \theta _{23}+\xi \right) }\sin \beta \right)
H_{r} \label{n1}
\end{equation}
and
\begin{equation}
\varphi _{2}+i\chi _{2}=\sum_r \left( q_{r1}e^{i\xi }\sin \beta
+q_{r2}e^{-i\theta _{23}}\cos \beta \right) H_{r}.  \label{n2}
\end{equation}

Then, we arrive to the final general form of the neutral Higgs boson
couplings for the up-type quarks:

\begin{equation}
\mathcal{L}_{up}^{neutral}=\overline{u}_{i}\left(S_{ijr}^{u}+%
\gamma^{5}P_{ijr}^{u}\right)
u_{j}H_{r}+\overline{u}_{i}M_{ij}^{U}u_{j}, \label{ygral}
\end{equation}

with

\begin{eqnarray}
S_{ijr}^{u} &=&\frac{1}{2v}M_{ij}^{U}\left( q_{k1}^{\ast
}+q_{k1}-\tan \beta \left( q_{k2}^{\ast }e^{i\left( \theta _{23}+\xi
\right) }+q_{k2}e^{-i\left(
\theta _{23}+\xi \right) }\right) \right)   \nonumber \\
&&+\frac{1}{2\sqrt{2}\cos \beta }\left( q_{k2}^{\ast }e^{i\theta _{23}}%
\widetilde{Y}_{2ij}^{U}+q_{k2}e^{-i\theta _{23}}\widetilde{Y}%
_{2ij}^{U\dagger }\right)   \label{sugral}
\end{eqnarray}

and

\begin{eqnarray}
P_{ijr}^{u} &=&\frac{1}{2v}M_{ij}^{U}\left( q_{k1}^{\ast
}-q_{k1}-\tan \beta \left( q_{k2}^{\ast }e^{i\left( \theta _{23}+\xi
\right) }-q_{k2}e^{-i\left(
\theta _{23}+\xi \right) }\right) \right)   \nonumber \\
&&+\frac{1}{2\sqrt{2}\cos \beta }\left( q_{k2}^{\ast }e^{i\theta _{23}}%
\widetilde{Y}_{2ij}^{U}-q_{k2}e^{-i\theta _{23}}\widetilde{Y}%
_{2ij}^{U\dagger }\right).   \label{pugral}
\end{eqnarray}

Similarly, for the down-type quarks we find:
\begin{equation}
\mathcal{L}_{down}^{neutral}=\overline{d}_{i}\left(
S_{ijr}^{d}+\gamma ^{5}P_{ijr}^{d}\right)
d_{j}H_{r}+\overline{d}_{i}M_{ij}^{D}d_{j},
\end{equation}

with
\begin{eqnarray}
S_{ijr}^{d} &=&\frac{1}{2v}M_{ij}^{D}\left[ q_{k1}+q_{k1}^{\ast
}-\tan \beta \left( q_{k2}^{\ast }e^{i\left( \theta _{23}+\xi
\right) }+q_{k2}e^{-i\left(
\theta _{23}+\xi \right) }\right) \right]   \nonumber \\
&&+\frac{1}{2\sqrt{2}\cos \beta }\left( q_{k2}e^{-i\theta
_{23}}Y_{2}^{D}+q_{k2}^{\ast }e^{i\theta
_{23}}\widetilde{Y}_{2}^{D\dagger }\right)   \label{sdgral}
\end{eqnarray}

and
\begin{eqnarray}
P_{ijr}^{d} &=&\frac{1}{2v}M_{ij}^{D}\left[ q_{k1}-q_{k1}^{\ast
}+\tan \beta \left( q_{k2}^{\ast }e^{i\left( \theta _{23}+\xi
\right) }-q_{k2}e^{-i\left(
\theta _{23}+\xi \right) }\right) \right]   \nonumber \\
&&+\frac{1}{2\sqrt{2}\cos \beta }\left( q_{k2}e^{-i\theta _{23}}\widetilde{Y}%
_{2}^{D}-q_{k2}^{\ast }e^{i\theta _{23}}\widetilde{Y}_{2}^{D\dagger
}\right). \label{pdgral}
\end{eqnarray}
On the other hand, the Yukawa couplings for charged states are given
by:
\begin{eqnarray}
\mathcal{L}_{Y}^{H^+} &=&\overline{u}\left[ \varphi _{1}^{+}V\left(
\frac{\sqrt{2}}{v_{1}}M^{D}-e^{i\xi }\tan{\beta}\widetilde{Y}%
_{2}^{D}\right) \frac{\mathbb{I}+\gamma ^{5}}{2}+\varphi _{2}^{+}V\widetilde{%
Y}_{2}^{D}\frac{\mathbb{I}+\gamma ^{5}}{2}\right.   \nonumber \\
&&\left. -\varphi _{1}^{+}\frac{\mathbb{I}-\gamma ^{5}}{2}\left( \frac{\sqrt{%
2}}{v_{1}}M^{U}-e^{i\xi }\tan{\beta}\widetilde{Y}_{2}^{U\dag
}\right) V-\varphi _{2}^{+}\frac{\mathbb{I}-\gamma ^{5}}{2}\widetilde{Y}%
_{2}^{U\dag }V\right] d  \nonumber \\
&&+h.c.
\end{eqnarray}

where $V$ denotes the CKM matrix. The physical eigenstates for the
charged Higgs boson $(H^+)$ can be obtain through the following
rotation:

\begin{equation}
\left(
\begin{array}{c}
\varphi _{1}^{\pm } \\
\varphi _{2}^{\pm }%
\end{array}%
\right) =\left(
\begin{array}{cc}
\cos \beta & -e^{\mp i\xi }\sin \beta \\
e^{\pm i\xi }\sin \beta & \cos \beta%
\end{array}%
\right) \left(
\begin{array}{c}
G^{\pm } \\
H^{\pm }%
\end{array}%
\right)
\end{equation}

Therefore, the Yukawa couplings for charged Higgs are
\begin{eqnarray}
\mathcal{L}_{Y}^{H^+} &=&\overline{u}\left[ H^{+}e^{-i\xi
}M^{U}V\frac{\mathbb{I} -\gamma ^{5}}{\sqrt{2}}-H^{+}e^{-i\xi
}VM^{D}\frac{
\mathbb{I}+\gamma ^{5}}{\sqrt{2}}\right.   \nonumber \\
&&\left. +\frac{1}{\cos \beta }H^{+}\left( V\widetilde{Y}_{2}^{D}\frac{%
\mathbb{I}+\gamma ^{5}}{2}-\widetilde{Y}_{2}^{U\dag }V\frac{\mathbb{I}%
-\gamma ^{5}}{2}\right) \right] d  \nonumber \\
&&+h.c.  \label{lch}
\end{eqnarray}%

\section{Some limiting cases}

\subsection{The THDM-V with explicit CP violation (2HDM-Va)}

In this case we assume the hermiticity condition for the Yukawa
matrices, but the Higgs sector could be CP violating. For simplicity
we shall consider that the Yukawa matrices obey a four-texture form,
and CP is violated explicitly in the Higgs sector.

As it is discussed in the appendix \ref{app1}, the assumption of
universal 4-textures for the Yukawa matrices, allows to express one
Yukawa matrix in terms of the quark masses, and parametrize the
FCNSI in terms of the unknown coefficients
$\chi _{ij}$, namely $\widetilde{Y}_{2ij}^{U}=\chi _{ij}\frac{\sqrt{%
m_{i}m_{j}}}{v}$, where the hermiticity condition reads $\chi _{ij}=\chi _{ij}^{\dag }$%
. These parameters can be constrained by considering all types of
low energy FCNC transitions. Although these constraints are quite
strong for transitions involving the first and second families, as
well as for the b-quark, it turns out that they are rather mild for
the top quark.

Then, from (\ref{sugral}) and (\ref{pugral}), one obtains within
2HDM-Va, the following expressions for the couplings of the neutral
Higgs bosons with up-type quarks, namely:

\begin{equation}
S_{ijr}^{u}=\frac{1}{2v}M_{ij}^{U}\left[ q_{r1}^{\ast }+q_{r1}-\tan
\beta
\left( q_{r2}^{\ast }+q_{r2}\right) \right] +\frac{\sqrt{m_{i}m_{j}}}{2\sqrt{%
2}v\cos \beta }\chi _{ij}\left( q_{r2}^{\ast }+q_{r2}\right)
\label{sva}
\end{equation}

and
\begin{equation}
P_{ijr}^{u}=\frac{1}{2v}M_{ij}^{U}\left[ q_{r1}^{\ast }-q_{r1}-\tan
\beta
\left( q_{r2}^{\ast }-q_{r2}\right) \right] +\frac{\sqrt{m_{i}m_{j}}}{2\sqrt{%
2}v\cos \beta }\chi _{ij}\left( q_{r2}^{\ast }-q_{r2}\right),
\label{pva}
\end{equation}

similar expressions can be obtained for the down-type quarks and
leptons, as well as for the charged Higgs couplings.

\subsection{The Yukawa Lagragian for the 2HDM-Vb}

In this case we shall consider that the Higgs sector is CP
conserving, while the Yukawa matrices could be non-hermitian. Then,
without loss of generality, we can assume that $H_{3}$ is CP odd,
while $H_{1}$ and $H_{2}$ are CP even. Then: $\cos \theta _{12}=\sin
\left( \beta -\alpha \right)$, $\sin \theta _{12}=\cos \left( \beta
-\alpha \right)$, $\sin \theta _{13}=0$, and $ e^{-i\theta
_{13}}=1$. The mixing angles $\alpha $ and $\beta $ that appear in
the neutral Higgs mixing, corresponds to the standard notation. The
expressions (\ref{phys-eigen}) for the neutral Higgs masses
eigenstates can be written now in terms of the angles $\alpha $ and
$\beta $:

\begin{eqnarray}
\Phi _{a}^{0} &=&\frac{1}{\sqrt{2}}\left( v+h^{0}\sin \left( \beta
-\alpha
\right) +H^{0}\cos \left( \beta -\alpha \right) +iG^{0}\right) \widehat{v}%
_{a}  \nonumber \\
&&+\frac{1}{\sqrt{2}}\left( h^{0}\cos \left( \beta -\alpha \right)
-H^{0}\sin \left( \beta -\alpha \right) +iA^{0}\right)
\widehat{w}_{a}, \label{higgscpc}
\end{eqnarray}

where $a=1,2$, and for the CP-conserving limit $\widehat{v}_{a}$\ and $\widehat{w%
}_{a}$ have a vanishing phase $\xi =0$. Additionally, when one
assumes a 4-texture for the Yukawa matrices, the Higgs-fermion
couplings further simplify as $\widetilde{Y}_{2ij}^{U}=\chi
_{ij}\frac{\sqrt{m_{i}m_{j}}}{v}$. Then, the corresponding
coefficient equation for up sector and $h^0$ ($r=1$) are

\begin{equation}
S_{ij1}^{u}=\frac{1}{v}M_{ij}^{U}\left[ \sin (\beta -\alpha )-\tan
\beta \cos (\beta -\alpha )\right]
+\frac{\sqrt{m_{i}m_{j}}}{2\sqrt{2}v}\frac{\cos (\beta -\alpha
)}{\cos \beta }\left( \chi _{ij}+\chi _{ij}^{\dag }\right)
\label{svb}
\end{equation}

\begin{equation}
P_{ij1}^{u}=\frac{\sqrt{m_{i}m_{j}}}{2\sqrt{2}v}\frac{\cos (\beta -\alpha )}{%
\cos \beta }\left( \chi _{ij}-\chi^{\dag }_{ij}\right) \label{pvb}
\end{equation}

For $H^0$ $(r=2)$ one finds:

\begin{equation}
S_{ij2}^{u}=\frac{1}{v}M_{ij}^{U}\left[ \cos (\beta -\alpha )-\tan
\beta \sin (\beta -\alpha )\right]
+\frac{\sqrt{m_{i}m_{j}}}{2\sqrt{2}v}\frac{\sin (\beta -\alpha
)}{\cos \beta }\left( \chi _{ij}+\chi _{ij}^{\dag }\right) ,
\end{equation}%
\begin{equation}
P_{ij2}^{u}=\frac{\sqrt{m_{i}m_{j}}}{2\sqrt{2}v}\frac{\sin (\beta -\alpha )}{%
\cos \beta }\left( \chi _{ij}-\chi _{ij}^{\dag }\right)
\end{equation}%
Finally, for $A^0$ $(r=3)$ one obtains:

\begin{equation}
S_{ij3}^{u}=i\frac{\sqrt{m_{i}m_{j}}}{2\sqrt{2}v\cos \beta }\left(
\chi _{ij}-\chi _{ij}^{\dag }\right) ,  \label{svb3}
\end{equation}%
\begin{equation}
P_{ij3}^{u}=\frac{i}{2v}M_{ij}^{U}\tan \beta -i\frac{\sqrt{m_{i}m_{j}}}{2%
\sqrt{2}v\cos \beta }\left( \chi _{ij}+\chi _{ij}^{\dag }\right)
\label{pvb3}
\end{equation}

\subsection{The 2HDM of type I, II and III}

It is interesting, and illustrative, to consider the limit when the
general model becomes the 2HDM-III, within 2HDM-III, one has that
the Yukawa matrices obey a 4-texture form, and also
$Y_{f}=Y_{f}^{\dagger }$, namely:
\[
\widetilde{Y}_{2ij}^{U}=\chi _{ij}\frac{\sqrt{m_{i}m_{j}}}{v}.
\]

The condition of Hermiticity means then $\chi _{ij}=\chi _{ij}^{\dag
}$. Within 2HDM III, we shall consider that the Higgs sector is CP
conserving. Therefore the Yukawa couplings take the following form.
For $h^0$ one gets,

\begin{eqnarray}
S_{ij1}^{u} &=&\frac{1}{v}M_{ij}^{U}\left( \sin (\beta -\alpha
)+\tan \beta
\cos (\beta -\alpha )\right)   \nonumber \\
&&-\frac{\chi _{ij}\sqrt{m_{i}m_{j}}}{\sqrt{2}v}\frac{\cos (\beta -\alpha )}{%
\cos \beta },
\end{eqnarray}%
while
\begin{equation}
P_{ij1}^{u}=0.
\end{equation}%
Then for $H^0$,

\begin{equation}
S_{ij2}^{u}=-\frac{1}{v}M_{ij}^{U}\frac{\sin \alpha }{\cos \beta }+\frac{%
\chi _{ij}\sqrt{m_{i}m_{j}}}{\sqrt{2}v}\frac{\sin (\beta -\alpha
)}{\cos \beta },
\end{equation}
and
\begin{equation}
P_{ij2}^{u}=0.
\end{equation}%
Then for $A^0$,

\begin{equation}
S_{ij3}^{u}=0,
\end{equation}
and
\begin{equation}
P_{ij3}^{u}=-i\frac{\chi _{ij}\sqrt{m_{i}m_{j}}}{\sqrt{2}v\cos \beta
}.
\end{equation}

\begin{center}
\begin{table}
\begin{tabular}{|c|c|c|c|}
\hline Model type & $H_1$  & $H_2$ & $H_3$\\ \hline
2HDM-I & $S\neq0$, $P=0$ & $S\neq0$, $P=0$ & $S=0$, $P\neq0$ \\
2HDM-II & $S\neq0$, $P=0$ & $S\neq0$, $P=0$ & $S=0$, $P\neq0$ \\
2HDM-III & $S\neq0$, $P=0$ & $S\neq0$, $P=0$ & $S=0$, $P\neq0$ \\
2HDM-V & $S$, $P\neq0$ & $S$, $P\neq0$ & $S$, $P\neq0$ \\ \hline
\end{tabular}
\caption{Yukawa couplings for neutral Higgs. The $S$ and $P$ can be
for up and down sector.} \label{table3}
\end{table}
\end{center}
%
One can also reduce the general model to the 2HDM types I and II, by
eliminating some of the Yukawa matrices $Y_2^{U,D}=0$ and
$Y_1^U=Y_2^D=0$, accordingly. The tables \ref{table3} and
\ref{table4} summarize the corresponding results, they include the
expressions for the neutral Higgs couplings with up and down type
quarks, and similar results hold for the leptons.

\begin{center}
\begin{table}
\begin{tabular}{|c|c|c|c|c|}
\hline r & $S^u_{ijr}$  & $P^u_{ijr}$ & $S^d_{ijr}$ & $P^d_{ijr}$\\
\hline
1 & $-\frac{\cos\alpha}{v\sin\beta}M_{ij}^U$ & $0$ & $-\frac{\cos\alpha}{v\sin\beta}M_{ij}^D$ & $0$ \\
2 & $-\frac{\sin\alpha}{v\sin\beta}M_{ij}^U$ & $0$ & $-\frac{\sin\alpha}{v\sin\beta}M_{ij}^D$ & $0$ \\
3 & $0$ & $\frac{i\cot\beta}{v}M_{ij}^U$ & $0$ & $-\frac{i\cot\beta}{v}M_{ij}^D$ \\
\hline
\end{tabular}
\caption{Explicit values of the Yukawa couplings for neutral Higgs
in 2HDM-I.} \label{table4}
\end{table}
\end{center}

\begin{center}
\begin{table}
\label{table5}
\begin{tabular}{|c|c|c|c|c|}
\hline r & $S^u_{ijr}$  & $P^u_{ijr}$ & $S^d_{ijr}$ & $P^d_{ijr}$\\
\hline
1 & $-\frac{\cos\alpha}{v\sin\beta}M_{ij}^U$ & $0$ & $\frac{\sin\alpha}{v\cos\beta}M_{ij}^D$ & $0$ \\
2 & $-\frac{\sin\alpha}{v\sin\beta}M_{ij}^U$ & $0$ & $-\frac{\cos\alpha}{v\cos\beta}M_{ij}^D$ & $0$ \\
3 & $0$ & $\frac{i\cot\beta}{v}M_{ij}^U$ & $0$ & $\frac{i\tan\beta}{v}M_{ij}^D$ \\
\hline
\end{tabular}
\caption{Explicit values of the Yukawa couplings for neutral Higgs
in 2HDM-II.}
\end{table}
\end{center}

\section{Probing the CP violating Higgs couplings through the decay $h\to
c\bar{b}W $}

\begin{figure}[tbp]
\centering
\includegraphics[scale=0.75]{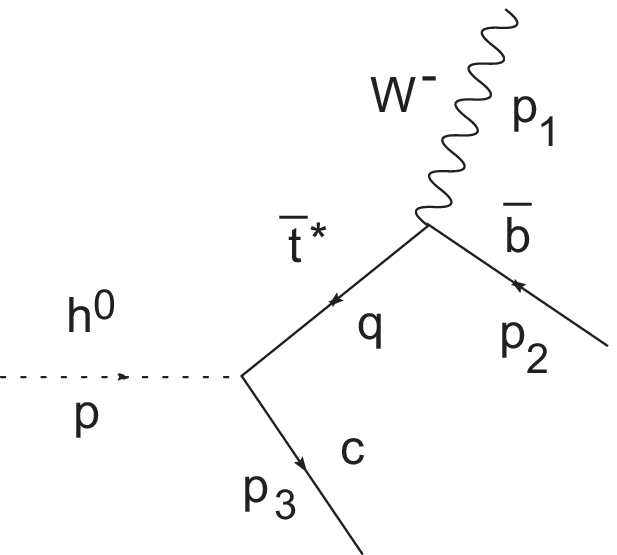}
\hspace{1cm}
\includegraphics[scale=0.75]{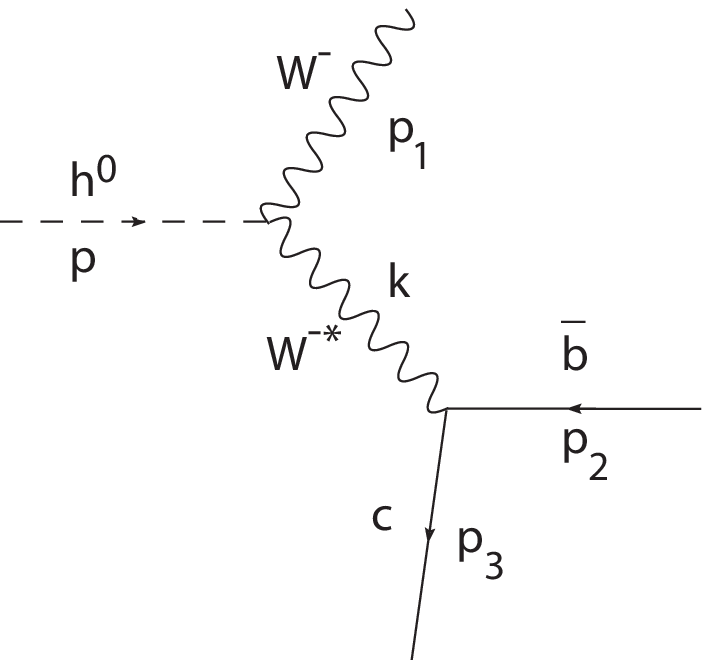}
\caption{Tree level Feynman diagrams for the decay. Right diagram is
for $h\longrightarrow W^{-}\overline{b}c$ while left diagram is for
$h\longrightarrow W^{-}\overline{b}c$.} \label{f1}
\end{figure}

In this section we shall evaluate the asymmetry coefficient for the
decay $h\to c\bar{b}W$ in order to analyze presence of both FCNSI
and CPV within the 2HDM-X. In the SM the FCNC are suppressed, but in
the 2HDM extensions these processes are found even at tree level. We
consider the neutral Higgs boson decay $h\longrightarrow
W\overline{b}c$\ at tree level. Two diagrams contribute to this
decay, the first one is through the FCNC coupling $h\longrightarrow
\overline{t}^{\ast }c\longrightarrow W^{-}\overline{b}c$, its
Feynman diagram is shown on left figure \ref{f1}. The other one is
through $h\longrightarrow W^{+\ast }W^{-}\longrightarrow
W^{-}\overline{b}c$, also shown in figure \ref{f1}.

The couplings of the neutral Higgs with the quarks and the charged
boson W with the neutral Higgs are written as $i\left( S_{231}^{u}+\gamma ^{5}P_{231}^{u}\right) $ and $%
igM_{W}q_{11}g^{\mu \nu },$ respectively. The other vertices are the
usual SM contribution. The average amplitude for these diagrams is
thus

\begin{equation}
\overline{\vert \mathcal{M}\vert }^{2}=\overline{\vert
\mathcal{M}_1\vert}^{2}+\overline{\vert
\mathcal{M}_2\vert}^{2}+\overline{ \mathcal{M}_{1}^{\dagger }\mathcal{M}_{2}}+\overline{%
\mathcal{M}_{2}^{\dagger}\mathcal{M}_{1}}
\end{equation}

We can obtain an approximation when the terms proportional to the
charm and bottom masses are neglected. Then, the expressions for the
squared amplitudes are
\begin{eqnarray}
\overline{\left\vert \mathcal{M}_{1}\right\vert }^{2} &=&\frac{g^{2}}{4M_{W}^{2}}%
\left\vert P_{t}(q)\right\vert ^{2}[4\left\vert
S_{231}^{u}-P_{231}^{u}\right\vert ^{2}p_{1}\cdot p_{2}p_{1}\cdot
qp_{3}\cdot q+2\left\vert S_{231}^{u}-P_{231}^{u}\right\vert
^{2}M_{W}^{2}p_{2}\cdot qp_{3}\cdot q  \nonumber \\
&&\left. +\left( \left\vert S_{231}^{u}+P_{231}^{u}\right\vert
^{2}m_{t}^{2}-\left\vert S_{231}^{u}-P_{231}^{u}\right\vert
^{2}q^{2}\right) \left( 2p_{1}\cdot p_{2}p_{1}\cdot
p_{3}+M_{W}^{2}p_{2}\cdot p_{3}\right) \right] ,  \label{m11f}
\end{eqnarray}

\begin{equation}
\overline{\left\vert \mathcal{M}_{2}\right\vert }^{2}=g^{4}\left(
q_{11}\right) ^{2}\left\vert V_{cb}\right\vert
^{2}|P_{W}(k)|^{2}\left( M_{W}^{2}p_{2}\cdot p_{3}+2p_{1}\cdot
p_{2}p_{1}\cdot p_{3}\right) ,  \label{m22f}
\end{equation}

\begin{equation}
\overline{\mathcal{M}_{1}^{\dagger
}\mathcal{M}_{2}}=\frac{g^{4}m_{t}}{M_{W}}\left( S_{231}^{u\ast
}+P_{231}^{u\ast }\right) q_{11}V_{cb}P_{t}^{\ast }(q)P_{W}(k)\left(
M_{W}^{2}p_{2}\cdot p_{3}+2p_{1}\cdot p_{2}p_{1}\cdot p_{3}\right)
\label{m1m2f}
\end{equation}%
and
\begin{equation}
\overline{\mathcal{M}_{2}^{\dagger
}\mathcal{M}_{1}}=\frac{g^{4}m_{t}}{M_{W}}\left(
S_{231}^{u}+P_{231}^{u}\right) q_{11}V_{cb}P_{W}^*(k)P_{t}(q)\left(
M_{W}^{2}p_{2}\cdot p_{3}+2p_{1}\cdot p_{2}p_{1}\cdot p_{3}\right)
.\label{m2m1f}
\end{equation}

where the W boson propagator is written in the Feynman-t'Hooft gauge
$P_{W}(k)=\left( k^{2}-M_{W}^{2}+iM_{W}\Gamma _{W}\right)^{-1}$ and
$P_{t}\left( q\right) =\left( q^{2}-m_{t}^{2}+im_{t}\Gamma
_{t}\right) ^{-1}$. In order to find the
asymmetry coefficient we also need to calculate the conjugate decay, that is, $%
h\longrightarrow W^{+}b\overline{c}$. We denote the average
amplitude as
\begin{equation}
\overline{\vert \widetilde{\mathcal{M}}\vert }^{2} =\overline{\vert \widetilde{\mathcal{M}_{1}}%
\vert}^{2} +\overline{\vert\widetilde{\mathcal{M}_{2}}\vert}^{2}+\overline{\widetilde{%
\mathcal{M}_{1}}^{\dagger}\widetilde{\mathcal{M}_{2}}} +\overline{\widetilde{\mathcal{M}_{2}}^{\dagger }%
\widetilde{\mathcal{M}_{1}}}.
\end{equation}

The square terms are the same as the above, $\overline{\left\vert \widetilde{%
\mathcal{M}}_{1,2}\right\vert }^{2}=\overline{\left\vert
\mathcal{M}_{1,2}\right\vert }^{2}$, while for the interference
terms we have

\begin{equation}
\overline{\widetilde{\mathcal{M}}_{1}^{\dagger }\widetilde{\mathcal{M}}_{2}}=\frac{g^{4}m_{t}}{%
M_{W}}\left( S_{231}^{u}+P_{231}^{u}\right)
q_{11}V_{cb}P_{t}(q)P_{W}^{\ast}(k)\left( M_{W}^{2}p_{2}\cdot
p_{3}+2p_{1}\cdot p_{2}p_{1}\cdot p_{3}\right)  \nonumber
\end{equation}%
and

\begin{equation}
\overline{\widetilde{\mathcal{M}}_{2}^{\dagger }\widetilde{\mathcal{M}}_{1}}=\frac{g^{4}m_{t}}{%
M_{W}}\left( S_{231}^{u\ast }+P_{231}^{u\ast }\right)
q_{11}V_{cb}P_{t}^{\ast}(q)P_{W}(k)\left( M_{W}^{2}p_{2}\cdot
p_{3}+2p_{1}\cdot p_{2}p_{1}\cdot p_{3}\right) .
\end{equation}
Then, the width for the decay is
\begin{equation}
\Gamma _{h\longrightarrow W\overline{b}c}=\frac{m_{h}}{256\pi
^{3}}\int
\int_{R_{xy}}\left(\overline{\left\vert \mathcal{M}_{1}\right\vert }^{2}+\overline{%
\left\vert \mathcal{M}_{2}\right\vert }^{2}+\overline{\mathcal{M}_{1}^{\dagger}\mathcal{M}_{2}}+\overline{%
\mathcal{M}_{2}^{\dagger }\mathcal{M}_{1}}\right)dxdy,
\end{equation}
where the dimensionless variables are defined as $x=\frac{2E_{1}}{m_{h}}$ and $y=\frac{2E_{2}}{%
m_{h}}$. All details about the decay kinematic were included in the
appendix \ref{app2}. The definition for the asymmetry coefficient is
\begin{equation}
A_{CPV}=\frac{\Gamma _{h\longrightarrow W^+\overline{b}c}-\Gamma
_{h\longrightarrow W^-b\overline{c}}}{\Gamma _{h\longrightarrow W^+\overline{b}%
c}+\Gamma _{h\longrightarrow W^-b\overline{c}}}.  \label{asy}
\end{equation}
The final result, for the decay asymmetry is given by:
\begin{equation}
A_{CPV}\left( S_{ijk}^{u},P_{ijk}^{u},\textrm{Re}\left(
q_{k1}\right) ,m_{h}\right) =\frac{2V_{cb}\textrm{Re}\left(
q_{k1}\right) \textrm{Im}\left( S_{ijk}^{u}+P_{ijk}^{u}\right)
\left( J_{10}+J_{12}\right) }{f\left(
S_{ijk}^{u},P_{ijk}^{u},\textrm{Re}\left( q_{k1}\right)
,m_{h}\right) }, \label{asymmetry}
\end{equation}%
where%
\begin{eqnarray}
f\left( S_{ijk}^{u},P_{ijk}^{u},\textrm{Re}\left( q_{k1}\right)
,m_{h}\right) &=&\frac{1}{4g}\left[ \left\vert
S_{ijk}^{u}-P_{ijk}^{u}\right\vert ^{2}\left(
J_{1}+J_{2}-J_{4}-J_{6}\right) +\left\vert
S_{ijk}^{u}+P_{ijk}^{u}\right\vert ^{2}\left( J_{3}+J_{5}\right)
\right]
\nonumber \\
&&+g\textrm{Re}\left( q_{k1}\right) ^{2}\left\vert V_{cb}\right\vert
^{2}\left( J_{7}+J_{8}\right) +2\textrm{Re}\left( q_{k1}\right)
\textrm{Re} \left( S_{ijk}^{u}+P_{ijk}^{u}\right) V_{cb}\left(
J_{9}+J_{11}\right) . \label{f-aux}
\end{eqnarray}
The $J$'s are integrals obtained from the decay kinematic, which are
shown in appendix \ref{app2} as well as the others parameters
defined in previous sections.

\subsection{Asymmetry in 2HDM-Va}

Let us discuss now the resulting expression for $A_{CPV}$ for two
subcases within 2HDM-V. We fix $i=2$ and $j=3$ in equations
(\ref{sva}) and (\ref{pva})in order to obtain the appropriate
parameters within 2HDM-Va, then we find:
\begin{equation}
S_{231}^{u}=\frac{\sqrt{m_{c}m_{t}}}{2\sqrt{2}v\cos \beta }\chi
_{23}\left( q_{12}^{\ast }+q_{12}\right)
\end{equation}

and
\begin{equation}
P_{231}^{u}=\frac{\sqrt{m_{c}m_{t}}}{2\sqrt{2}v\cos \beta }\chi
_{23}\left( q_{12}^{\ast }-q_{12}\right) .
\end{equation}%
Then, the asymmetry coefficient is

\begin{equation}
A_{2HDM-Va}=\frac{\sqrt{m_{c}m_{t}}V_{cb}\chi _{23}\left(
J_{10}+J_{12}\right)
\cos^2 \theta _{12}\cos \theta _{13}\sin \theta _{13}}{\sqrt{2%
}M_{W}f\left( \beta ,\theta _{12},\theta _{13},\chi
_{23},m_{h}\right) \cos \beta },
\end{equation}%
where
\begin{eqnarray}
f\left( \beta ,\theta _{12},\theta _{13},\chi _{23},m_{h}\right)  &=&\frac{%
m_{c}m_{t}\chi _{23}^{2}}{32M_{W}^{2}\cos ^{2}\beta }\left( \sin
^{2}\theta _{12}+\cos ^{2}\theta _{12}\sin ^{2}\theta _{13}\right)
\left(
J_{1}+J_{2}+J_{3}-J_{4}+J_{5}-J_{6}\right)   \nonumber \\
&&+\left\vert V_{cb}\right\vert ^{2}\cos ^{2}\theta _{12}\cos
^{2}\theta
_{13}\left( J_{7}+J_{8}\right) -\frac{\sqrt{m_{c}m_{t}}V_{cb}\chi _{23}}{%
\sqrt{2}M_{W}\cos \beta }\cos \theta _{12}\cos \theta _{13}\sin
\theta _{12}\left( J_{9}+J_{11}\right) .
\end{eqnarray}

\subsection{Asymmetry in 2HDM-Vb}

The appendix \ref{app1} shows the four-texture structure for Yukawa
matrices, nevertheless for practical evaluation of the asymmetry we
write the texture parameter in Euler complex form as
$\chi_{23}=|\chi_{23}|e^{i\nu_{23}}$. Then, we fix the equations
(\ref{svb}) and (\ref{pvb}) for $i=2$ and $j=3$ in order to obtain
the required element for 2HDM-Vb,
\begin{equation}
S_{231}^{u}=\frac{\cos (\beta -\alpha
)\sqrt{m_{c}m_{b}}}{2\sqrt{2}v\cos \beta }\left( \chi _{23}+\chi
_{23}^{\dagger }\right)   \label{s321b}
\end{equation}
and
\begin{equation}
P_{231}^{u}=\frac{\cos (\beta -\alpha
)\sqrt{m_{c}m_{b}}}{2\sqrt{2}v\cos \beta }\left( \chi _{23}-\chi
_{23}^{\dagger }\right).   \label{p231b}
\end{equation}
Then, for this case the asymmetry coefficient is given by:
\begin{equation}
A_{2HDM-Vb}=\frac{gV_{cb}\sqrt{m_{c}m_{t}}\left\vert \chi
_{23}\right\vert \sin
\nu _{23}\cos (\beta -\alpha )\sin \left( \beta -\alpha \right) }{\sqrt{2}%
M_{W}\cos \beta f\left( \alpha ,\beta ,\chi _{23},m_{h}\right)
}\left( J_{10}+J_{12}\right), \label{asymmetryb}
\end{equation}
where
\begin{eqnarray}
f\left( \alpha ,\beta ,\chi _{23},m_{h}\right)  &=&\frac{gm_{c}m_{t}}{%
32M_{W}^{2}}\frac{\cos ^{2}(\beta -\alpha )}{\cos ^{2}\beta
}\left\vert \chi _{23}\right\vert ^{2}\left(
J_{1}+J_{2}-J_{4}-J_{6}+J_{3}+J_{5}\right)
\nonumber \\
&&+\frac{g\sqrt{m_{c}m_{t}}V_{cb}}{\sqrt{2}M_{w}}\left\vert \chi
_{23}\right\vert \frac{\cos \nu _{23}\sin \left( \beta -\alpha
\right) \cos
(\beta -\alpha )}{\cos \beta }\left( J_{9}+J_{11}\right)   \nonumber \\
&&+g\sin ^{2}\left( \beta -\alpha \right) \left\vert
V_{cb}\right\vert ^{2}\left( J_{7}+J_{8}\right).   \label{f-auxb}
\end{eqnarray}

\subsection{Numerical results}
\begin{figure}
  \includegraphics[scale=0.75]{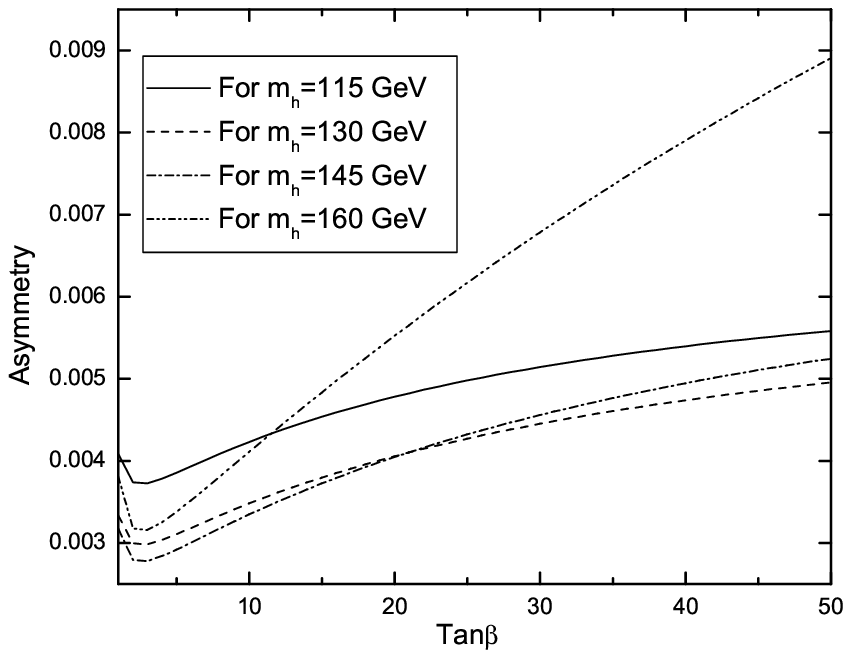}
  \includegraphics[scale=0.75]{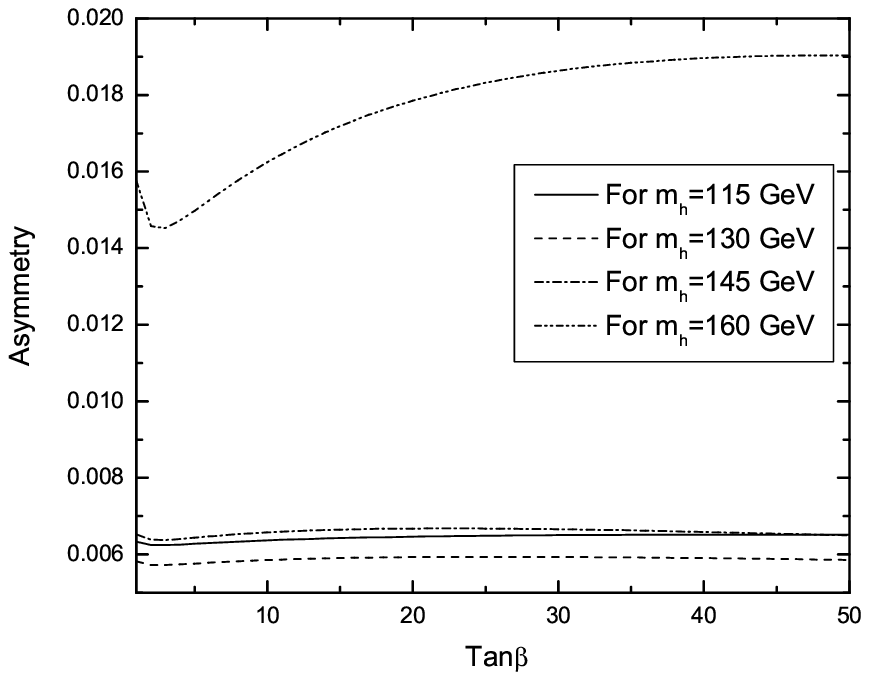}\\
  \caption{The asymmetry as function of $\tan\beta$ for scenario 1, on left for $|\chi_{23}|=0.1$
   and on right for $|\chi_{23}|=0.9$ }
   \label{scenario1}
\end{figure}

\begin{figure}
  \includegraphics[scale=0.75]{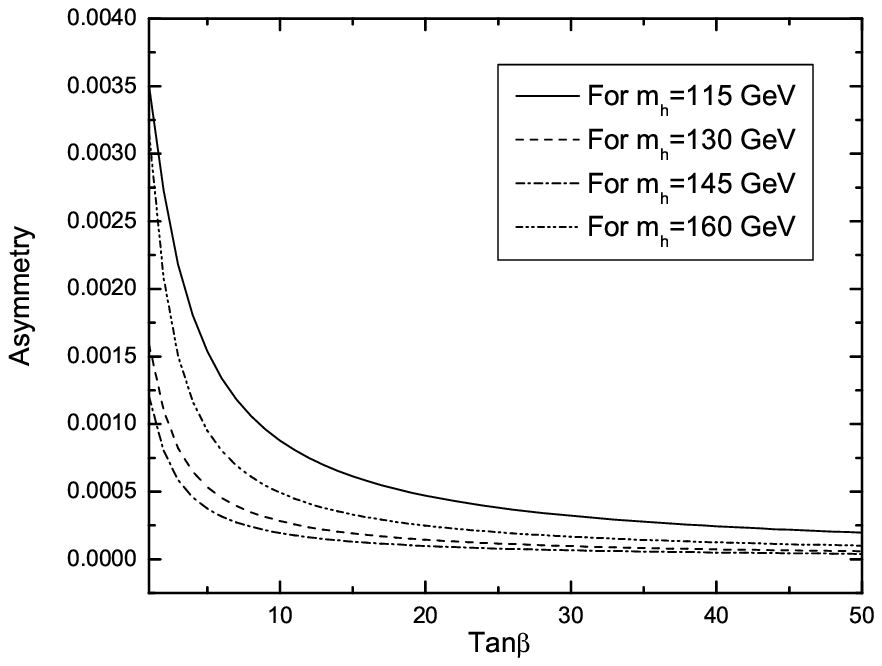}
  \includegraphics[scale=0.75]{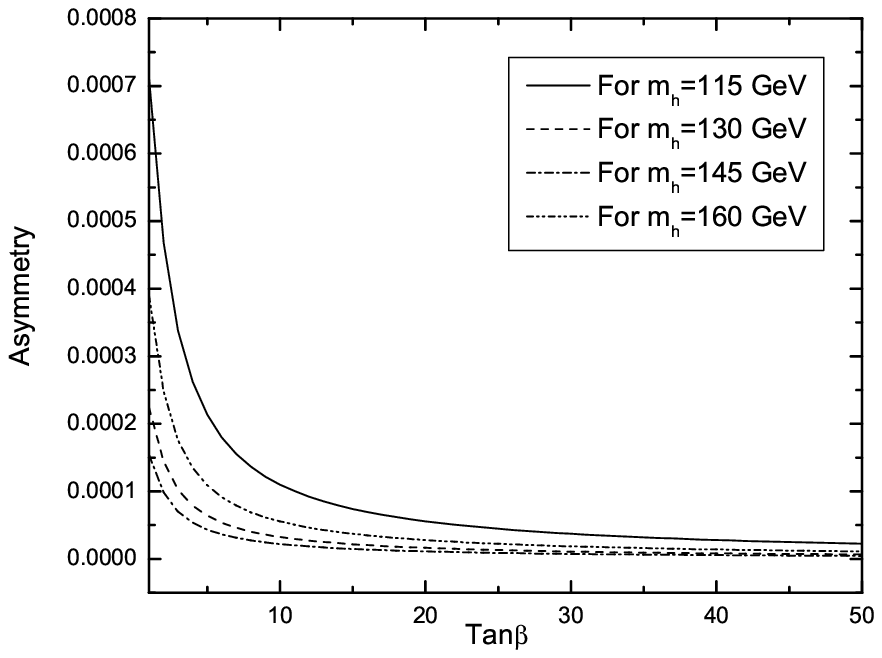}\\
  \caption{The asymmetry as function of $\tan\beta$ for scenario 2, on left for $|\chi_{23}|=0.1$
   and on right for $|\chi_{23}|=0.9$ }
   \label{scenario2}
\end{figure}

\begin{figure}
  \includegraphics[scale=0.75]{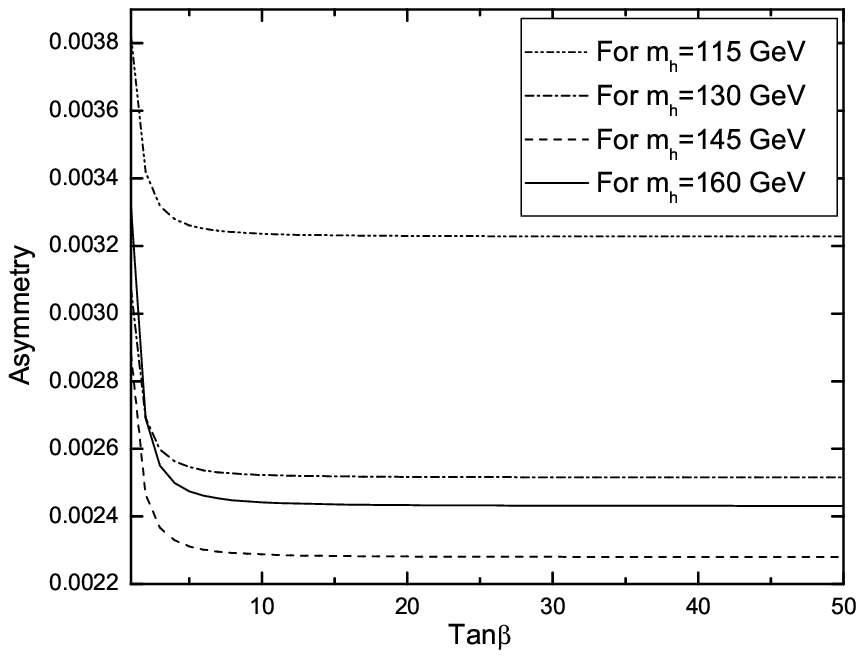}
  \includegraphics[scale=0.75]{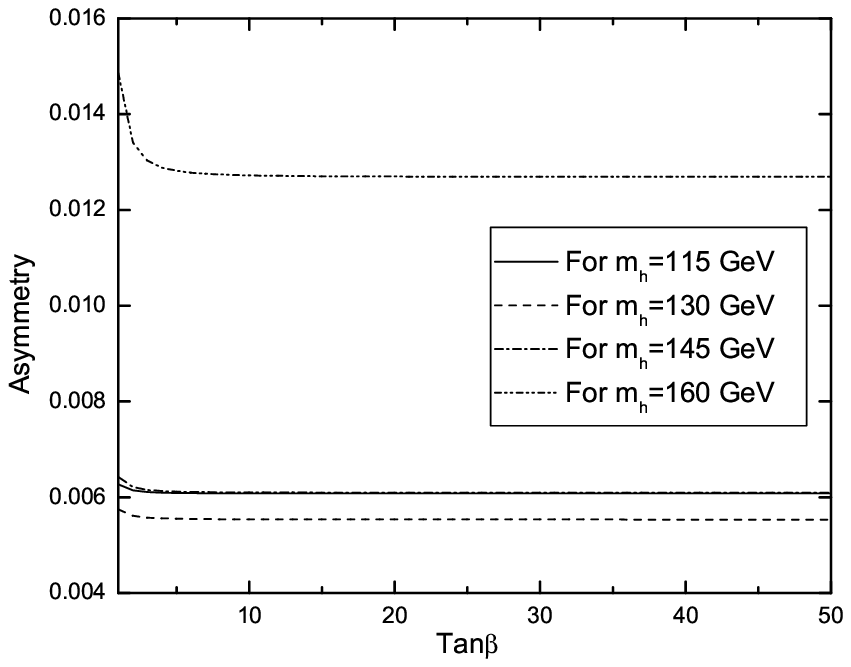}\\
  \caption{The asymmetry as function of $\tan\beta$ for scenario 3, on left for $|\chi_{23}|=0.1$
   and on right for $|\chi_{23}|=0.9$ }
   \label{scenario3}
\end{figure}

We shall discuss in detail the result for 2HDM-Vb. The asymmetry
depends of the five free parameters. One of them is the Higgs boson
mass which appears in the $J$ integrals, the other ones are the
mixing angles $\alpha$, $\beta$ and the complex parameter
$\chi_{23}$. The mixing angle $\beta$ is taken within the values
$1<\tan\beta<50$ \cite{pdg}. For the mixing angle $\alpha$ we study
three scenarios, $\alpha < \beta$, $\alpha \approx \beta$ and
$\alpha
> \beta$. The phase $\nu_{23}$ is fixed to the value $0.1$ in order to analyzed a similar value to the phase
from the CKM matrix. For each scenario we take two possible values
for $|\chi_{23}|$. Therefore, the scenarios studied here are:
\begin{enumerate}[i)]
  \item $\alpha < \beta$ for $|\chi_{23}|=0.9$ and
  $|\chi_{23}|=0.1$, figure \ref{scenario1}
  \item $\alpha \approx \beta$ for $|\chi_{23}|=0.9$ and
  $|\chi_{23}|=0.1$, figure \ref{scenario2}.
  \item $\alpha > \beta$ for $|\chi_{23}|=0.9$ and
  $|\chi_{23}|=0.1$, figure \ref{scenario3}.
\end{enumerate}

We use the reported values $m_t=171.2$ GeV, $m_b=4.2$ GeV,
$m_c=1.27$ GeV, $M_W=80.39$ GeV and $\sin\theta_W=0.231$ \cite{pdg}.

From figs \ref{scenario1}, \ref{scenario2} and \ref{scenario3} we
obtain asymmetry values of the order $10^{-3}$ to $10^{-2}$
($10^{-4}$ to $10^{-3}$ and $10^{-3}$) for scenario i) ( ii) and
iii))within 2HDM-Vb. We have also analyzed the numerical results for
the CP asummetry for case 2HDM-Va. We also find that the size of
this asymmetry depens strongly on the phases.

\section{Conclusions}

In this paper we have present a broad discussion of the most general
formulation of the Two-Higgs doublet extension of the SM, which we
name as 2HDM-X. Then, we have defined in a model named 2HDM-V, which
has the possibility of including both FCNC and CPV, and have
presented the corresponding Lagrangian for both the neutral and
charged Higgs sectors.

The limits when 2HDM-X reduces to one of the known versions (2HDM-I,
II, III) has also been discussed; in these cases each pattern of
Higgs-Yukawa couplings holds for all families. To identify the class
of family non-universal models, we have used the label 2HDM-IV,
where we include models where one Higgs doublet couples only to a
certain type of fermion, for instance to the top quark or the third
family, or to neutrinos only.

Finally, we have also evaluated the CPV asymmetry for the decay
$h\rightarrow c\bar{b}W$, which allows to test the presence of both
FCNC and CPV that associated with model V. We found that for certain
optimal range of parameters the decay asymmetry could be of
$O(10^{-2})$ to $O(10^{-4}$. These asymmetry values for three
scenarios were obtained in the case of the 2HDM-Vb. Similar results
arise within 2HDM-Va. The asymmetry behavior has a dependency
proportional to the mixing complex parameter $\chi_{23}$. The mixing
angles $\alpha$ and $\beta$ control the shape of the graphs. The
asymmetry keeps same shape for the Higgs boson mass range between
$115 $ GeV and $160$ GeV.

In order to detect this asymmetry we could have to resort to a
linear collider, since the final state seems difficult to
reconstruct at a hadron collider. Although a final conclusion would
require a detailed simulation study, which we plan to address in a
future publication \cite{working-progress}.

\noindent\textbf{Acknowledgments.}

\noindent We would like to thank Sistema Nacional de Investigadores
(Mexico) and CONACYT (Mexico).


\appendix
\section{2HDM-III with four-Textures}\label{app1}

Here we shall summarize the result for 2HDM-III, namely we assume
that both Yukawa matrices $Y^q_1$ and $Y^q_2$ have the four-texture
form and are Hermitic; following the conventions of
\cite{DiazCruz:2004tr}, the quark mass matrix is then written as:

\begin{displaymath}
M_q= \left( \begin{array}{ccc}
0 & C_{q} & 0 \\
C_{q}^{*} & \tilde{B}_{q} & B_{q} \\
0 & B_{q}^{*} & A_{q}
\end{array}\right).
\end{displaymath}
when $\tilde{B}_{q}\to 0$ one recovers the six-texture form.
We also consider the hierarchy: \\
$\mid A_{q}\mid \, \gg \, \mid \tilde{B}_{q}\mid,\mid B_{q}\mid
,\mid C_{q}\mid$, which is supported by the observed fermion masses.

Because of the hermicity condition, both $\tilde{B}_{q}$ and $A_{q}$
are real parameters, while the phases of $C_q$ and $B_q$,
$\Phi_{B_q,C_q}$, can be removed from the mass matrix $M_q$ by
defining: $M_q=P_q^\dagger \tilde{M}_q P_q$, where $P_q=diag[1,
e^{i\Phi_{C_q}},  e^{i(\Phi_{B_q}+\Phi_{C_q})}]$, and the mass
matrix $\tilde{M}_q$ includes only the real parts of $M_q$. The
diagonalization of $\tilde{M}_q$ is then obtained by an orthogonal
matrix $O_q$, such that the diagonal mass matrix is: $\bar{M}_{q} =
O_q^{T}\tilde{M}_{q}O_q$.


The lagrangian (2) can be expanded in terms of the mass-eigenstates
for the neutral ($h^0,H^0,A^0$) and charged Higgs bosons ($H^\pm$).
The interactions of the neutral Higgs bosons with the d-type and
u-type are given by ($u,u'=u,c,t.$ and $d,d\,'=d,s,b.$),

\begin{eqnarray}
{\cal{L}}_Y^{q} & = & \frac{g}{2}\left(\frac{m_d}{m_W}\right)
\bar{d}\left[\frac{ \, \cos\alpha}{\cos\beta}\delta_{dd'}+
\frac{\sqrt{2} \, \sin(\alpha - \beta)}{g \, \cos\beta}
\left(\frac{m_W}{m_d}\right)(\tilde{Y}_2^d)_{dd'}\right]d\,'H^{0}
\nonumber \\
                 &  &+ \frac{g}{2}\left(\frac{m_d}{m_W}\right)\bar{d}
\left[-\frac{\sin\alpha}{\cos\beta} \delta_{dd'}+ \frac{\sqrt{2} \,
\cos(\alpha - \beta)}{g \, \cos\beta}
\left(\frac{m_W}{m_d}\right)(\tilde{Y}_2^d)_{dd'}\right]d\,' h^{0}
\nonumber \\
                 & &+ \frac{ig}{2}\left(\frac{m_d}{m_W}\right)\bar{d}
\left[-\tan\beta \delta_{dd'}+  \frac{\sqrt{2} }{g \, \cos\beta}
\left(\frac{m_W}{m_d}\right)(\tilde{Y}_2^d)_{dd'}\right]
\gamma^{5}}d\,' A^{0 \nonumber \\
                 & &+ \frac{g}{2}\left(\frac{m_u}{m_W}\right)
\bar{u}\left[\frac{ \, \sin\alpha}{\sin\beta}\delta_{uu'}-
\frac{\sqrt{2} \, \sin(\alpha - \beta)}{g \, \sin\beta}
\left(\frac{m_W}{m_u}\right)(\tilde{Y}_2^u)_{uu'}\right]u'H^{0}
\nonumber \\
                 &  &+ \frac{g}{2}\left(\frac{m_u}{m_W}\right)\bar{u}
\left[\frac{\cos\alpha}{\sin\beta} \delta_{uu'}- \frac{\sqrt{2} \,
\cos(\alpha - \beta)}{g \, \sin\beta}
\left(\frac{m_W}{m_u}\right)(\tilde{Y}_2^u)_{uu'}\right]u' h^{0}
\nonumber \\
                 & &+ \frac{ig}{2}\left(\frac{m_u}{m_W}\right)\bar{u}
\left[-\cot\beta \delta_{uu'} + \frac{\sqrt{2} }{g \, \sin\beta}
\left(\frac{m_W}{m_u}\right)(\tilde{Y}_2^u)_{uu'}\right]
\gamma^{5}}u' A^{0.
\end{eqnarray}
The first term, proportional to $\delta_{qq'}$ corresponds to the
modification of the 2HDM-II over the SM result, while the term
proportional to $\tilde{Y}_2^q$ denotes the new contribution from
2HDM-III. Thus, the $f f' \phi^0$ couplings respect CP-invariance,
despite the fact that the Yukawa matrices include complex phases;
this follows because of the Hermiticity conditions imposed on both
$Y_1^q$ and $Y_2^q$.

The corrections to the quark flavor conserving (FC) and flavor
violating (FV) couplings, depend on the rotated matrix:
$\tilde{Y}_{2}^{q} = O_q^{T}P_qY_{2}^{q}P_q^\dagger O_q$. We will
evaluate $\tilde{Y}_{2}^{q}$ assuming that $Y_2^q$ has a
four-texture form, namely:

\begin{equation}
Y_{2}^{q}  = \left( \begin{array}{ccc}
0 & C_2^q & 0 \\
C_2^{q*} & \tilde{B}_2^q & B_2^q \\
0 & B_2^{q*} & A_2^q
\end{array}\right), \qquad
\mid A_2^q\mid \, \gg \, \mid \tilde{B}_2^q\mid,\mid B_2^q\mid ,\mid
C_2^q\mid.
\end{equation}

The matrix that diagonalizes the real matrix $\tilde{M}_{q}$ with
the four-texture form, is given by:

\begin{displaymath}
O_q = \left( \begin{array}{ccc}
\sqrt{\frac{\lambda^q_{2}\lambda^q_{3}(A_q-\lambda^q_{1})}{A_q(\lambda^q_{2}-\lambda^q_{1})
(\lambda^q_{3}-\lambda^q_{1})}}& \eta_q
\sqrt{\frac{\lambda^q_{1}\lambda^q_{3}
(\lambda^q_{2}-A_q)}{A_q(\lambda^q_{2}-\lambda^q_{1})(\lambda^q_{3}-\lambda^q_{2})}}
&
\sqrt{\frac{\lambda^q_{1}\lambda^q_{2}(A_q-\lambda^q_{3})}{A_q(\lambda^q_{3}-
\lambda^q_{1})(\lambda^q_{3}-\lambda^q_{2})}} \\
-\eta_q
\sqrt{\frac{\lambda^q_{1}(\lambda^q_{1}-A_q)}{(\lambda^q_{2}-\lambda^q_{1})
(\lambda^q_{3}-\lambda^q_{1})}} &
\sqrt{\frac{\lambda^q_{2}(A_q-\lambda^q_{2})}
{(\lambda^q_{2}-\lambda^q_{1})(\lambda^q_{3}-\lambda^q_{2})}} &
\sqrt{
\frac{\lambda^q_{3}(\lambda^q_{3}-A_q)}{(\lambda^q_{3}-\lambda^q_{1})(\lambda^q_{3}-
\lambda^q_{2})}} \\
\eta_q
\sqrt{\frac{\lambda^q_{1}(A_q-\lambda^q_{2})(A_q-\lambda^q_{3})}{A_q(\lambda^q_{2}
-\lambda^q_{1})(\lambda^q_{3}-\lambda^q_{1})}} &
-\sqrt{\frac{\lambda^q_{2}(A_q
-\lambda^q_{1})(\lambda^q_{3}-A_q)}{A_q(\lambda^q_{2}-\lambda^q_{1})(\lambda^q_{3}
-\lambda^q_{2})}} &
\sqrt{\frac{\lambda^q_{3}(A_q-\lambda^q_{1})(A_q-\lambda^q_{2})}
{A_q(\lambda^q_{3}-\lambda^q_{1})(\lambda^q_{3}-\lambda^q_{2})}}
\end{array}\right),
\end{displaymath}
where $m^q_1 = \mid \lambda^q _1\mid$, $m^q_2 = \mid \lambda^q
_2\mid$, $m^q_3 = \mid \lambda^q _3\mid$, and $\eta_q = \lambda^q_2/
m^q_2$ $(q=u,d)$. With $m_u= m^u_1$, $m_c= m^u_2$, and $m_t= m^u_3$;
$m_d= m^d_1$, $m_s= m^d_2$,
 and $m_b= m^d_3$.

Then the rotated form $\tilde {Y}_2^q$ has the general form,

\begin{eqnarray}
\tilde {Y}_2^q  & = & O_q^TP_qY_{2}^qP_q^{\dagger}O_q \nonumber \\
& = &\left( \begin{array}{ccc}
(\tilde {Y}_2^q)_{11}   & (\tilde {Y}_2^q)_{12}   & (\tilde {Y}_2^q)_{13}   \\
(\tilde {Y}_2^q)_{21}   & (\tilde {Y}_2^q)_{22}   & (\tilde {Y}_2^q)_{23}  \\
(\tilde {Y}_2^q)_{31}   & (\tilde {Y}_2^q)_{32}   & (\tilde
{Y}_2^q)_{33}
\end{array}\right).
\end{eqnarray}

However, the full expressions for the resulting elements have a
complicated form, as it can be appreciated, for instance, by looking
at the element $(\tilde{Y}_{2}^q)_{22}$, which is displayed here:

\begin{eqnarray}
(\tilde{Y}_2^q)_{22} &=& \eta_q [C^{q*}_2 e^{i\Phi_{C_q}} +C^q_2
e^{-i\Phi_{C_q}}] \frac{(A_q-\lambda^q_{2})}{m^q_3-\lambda^q_2 }
\sqrt{\frac{m^q_1 m^q_3 }{A_q m^q_2}} +
 \tilde{B}^q_2 \frac{A_q-\lambda^q_2}{ m^q_3-\lambda^q_2 }\nonumber \\
& & + A^q_2 \frac{A_q-\lambda^q_2}{ m^q_3-\lambda^q_2 } - [B^{q*}_2
e^{i\Phi_{B_q}} + B^q_2 e^{-i\Phi_{B_q}}]
\sqrt{\frac{(A_q-\lambda^q_{2})(m^q_3-A_q) } {m^q_3- \lambda^q_2}}
\end{eqnarray}
where we have taken the limits: $|A_q|, m^q_3, m^q_2 \gg m^q_1$. The
free-parameters are: $\tilde{B^q_{2}}, B^q_{2}, A^q_{2}, A_q$.

To derive a better suited approximation, we will consider the
elements of the Yukawa matrix $Y_2^l$ as having the same hierarchy
as the full mass matrix, namely:

\begin{eqnarray}
C^q_{2} & = &  c^q_{2}\sqrt{\frac{m^q_{1}m^q_{2}m^q_{3}}{A_q}}  \\
B^q_{2} & = &  b^q_{2}\sqrt{(A_q - \lambda^q_{2})(m^q_{3}-A_q)}  \\
\tilde{B}^q_{2} & = & \tilde{b}^q_{2}(m^q_{3}-A_q + \lambda^q_{2})  \\
A^q_{2} & = & a^q_{2}A_q.
\end{eqnarray}

Then, in order to keep the same hierarchy for the elements of the
mass matrix, we find that $A_q$ must fall within the interval $
(m^q_3- m^q_2) \leq A_q \leq m^q_3$. Thus, we propose the following
relation for $A_q$:

\begin{equation}
A_q  = m^q_{3}(1 -\beta_q z_q),
\end{equation}
where $z_q = m^q_{2}/m^q_{3} \ll 1$  and $0 \leq \beta_q \leq 1$.

Then, we introduce the matrix $\tilde{\chi}^q$ as follows:

\begin{eqnarray}
\left( \tilde {Y}_2^q \right)_{ij}
&=& \frac{\sqrt{m^q_i m^q_j}}{v} \, \tilde{\chi}^q_{ij} \nonumber\\
&=&\frac{\sqrt{m^q_i m^q_j}}{v}\, {\chi}^q_{ij} \, e^{i
\vartheta^q_{ij}}
\end{eqnarray}
which differs from the usual Cheng-Sher ansatz not only because of
the appearance of the complex phases, but also in the form of the
real parts ${\chi}^q_{ij} = |\tilde{\chi}^q_{ij}|$.

Expanding in powers of $z_q$, one finds that the elements of the
matrix $\tilde{\chi}^q$ have the following general expressions:

\begin{eqnarray}
\tilde{\chi}^q_{11} & = & [\tilde{b}^q_2-(c^{q*}_2e^{i\Phi_{C_q}}
+c^q_2e^{-i\Phi_{C_q}} )]\eta_q
    +[a^q_2+\tilde{b}^q_2-(b^{q*}_2e^{i\Phi_{B_q}} + b^q_2e^{-i\Phi_{B_q}} )]
         \beta_q \nonumber \\
\tilde{\chi}^q_{12} & = & (c^q_2e^{-i\Phi_{C_q}}-\tilde{b}^q_2)
-\eta_q[a^q_2+ \tilde{b}^q_2-(b^{q*}_2e^{i\Phi_{B_q}} +
b^q_2e^{-i\Phi_{B_q}} )] \beta_q
\nonumber \\
\tilde{\chi}^q_{13} & = & (a^q_2-b^q_2e^{-i\Phi_{B_q}}) \eta_q
\sqrt{\beta_q}
                           \nonumber  \\
\tilde{\chi}^q_{22}  & = & \tilde{b}^q_2\eta_q
+[a^q_2+\tilde{b}^q_2-(b^{q*}_2e^{i\Phi_{B_q}}
+b^q_2e^{-i\Phi_{B_q}} )]
         \beta_q \nonumber \\
\tilde{\chi}^q_{23} & = & (b^q_2e^{-i\Phi_{B_q}}-a^q_2)
                              \sqrt{\beta_q} \nonumber  \\
\tilde{\chi}^q_{33} & = & a^q_2
\end{eqnarray}

While the diagonal elements $\tilde{\chi}^q_{ii}$ are real, we
notice (Eqs. 14) the appearance of the phases in the off-diagonal
elements, which are essentially unconstrained by present low-energy
phenomena. As we will see next, these phases modify the pattern of
flavor violation in the Higgs sector. For instance, while the
Cheng-Sher ansatz predicts that the FCNC couplings
$(\tilde{Y}_2^q)_{13}$ and $(\tilde{Y}_2^q)_{23}$ vanish when $a_2^q
= b_2^q$, in our case this is no longer valid for $\cos\Phi_{B_q}
\neq 1$. Furthermore the FCNC couplings satisfy several relations,
such as: $|\tilde{\chi}^q_{23}| = |\tilde{\chi}^q_{13}|$, which
simplifies the parameter analysis. Similar expressions can be
obtained for the lepton sector.

\section{Decay kinematics for $h\rightarrow c\bar{b}W$}
\label{app2}

For sake of simplicity we introduce the dimensionless scaled
variables

\begin{equation}
\mu _{i}=\frac{m_{i}^{2}}{m_{h}^{2}}  \label{mu}
\end{equation}%
and

\begin{equation}
\left(
\begin{array}{ccc}
x, & y, & z%
\end{array}%
\right) =\left(
\begin{array}{ccc}
\frac{2E_{1}}{m_{h}}, & \frac{2E_{2}}{m_{h}}, & \frac{2E_{3}}{m_{h}}%
\end{array}%
\right) .  \label{xyz}
\end{equation}
With this notation we can write the energy conservation as

\begin{equation}
x+y+z=2.  \label{ce2}
\end{equation}

In the Higgs rest frame, we just consider the contribution of the $%
\mu _{1}$, because $\mu _{1}>>\mu _{2},\mu _{3}$, and find the
momentum expressions

\begin{eqnarray}
p_{1}\cdot p_{2} &=& \frac{m_{h}^{2}}{2}\left( x+y+\mu _{1}-1\right) , \\
p_{1}\cdot p_{3} &=& \frac{m_{h}^{2}}{2}\left( 1-y-\mu _{1}\right) , \\
p_{2}\cdot p_{3} &=& \frac{m_{h}^{2}}{2}\left( 1-x+\mu _{1}\right) , \\
p_{1}\cdot q &=& \frac{m_{h}^{2}}{2}\left( x+y+\mu _{1}-1\right) , \\
p_{2}\cdot q &=& \frac{m_{h}^{2}}{2}\left( x+y-\mu _{1}-1\right) , \\
p_{3}\cdot q &=& \frac{m_{h}^{2}}{2}\left( 2-x-y\right) , \\
k^{2} &=& m_{h}^{2}\left( 1+\mu _{1}-x\right) \\
q^{2} &=& m_{h}^{2}\left(x+y-1\right).  \label{ma}
\end{eqnarray}

Now, the functions $\left\vert P_{t}(q)\right\vert ^{2}$,
$\left\vert P_{W}(q)\right\vert ^{2}$, $P_{t}^{\ast }(q)P_{W}(k)$\
and $P_{t}(q)P_{W}^*(k)$ with the dimensionless variables can be
written as

\begin{equation}
\left\vert P_{t}(q)\right\vert
^{2}=\frac{1}{m_{h}^{4}}\frac{1}{\left( x+y-1-\mu \right) ^{2}+\mu
\Gamma ^{2}},  \label{pt2dls}
\end{equation}

\begin{equation}
\left\vert P_{W}(k)\right\vert
^{2}=\frac{1}{m_{h}^{4}}\frac{1}{\left( 1-x\right) ^{2}+\mu
_{1}\gamma ^{2}},  \label{pw2}
\end{equation}
\begin{equation}
P_{t}^{\ast }(q)P_{W}(k)=\frac{\left( x+y-1-\mu \right) \left( 1-x\right) +%
\sqrt{\mu \mu _{1}}\gamma \Gamma +i\left[ \sqrt{\mu }\Gamma \left(
1-x\right) -\sqrt{\mu _{1}}\gamma \left( x+y-1-\mu \right) \right] }{%
m_{h}^{4}\left[ \left( x+y-1-\mu \right) ^{2}+\mu \Gamma ^{2}\right]
\left[ \left( 1-x\right) ^{2}+\mu _{1}\gamma ^{2}\right] },
\label{pt*pw}
\end{equation}
and%
\begin{equation}
P_{t}(q)P_{W}^{\ast }(q)=\frac{\left( x+y-1-\mu \right) \left( 1-x\right) +%
\sqrt{\mu \mu _{1}}\gamma \Gamma -i\left[ \sqrt{\mu }\Gamma \left(
1-x\right) -\sqrt{\mu _{1}}\gamma \left( x+y-1-\mu \right) \right] }{%
m_{h}^{4}\left[ \left( x+y-1-\mu \right) ^{2}+\mu \Gamma ^{2}\right]
\left[ \left( 1-x\right) ^{2}+\mu _{1}\gamma ^{2}\right] },
\label{ptpw*}
\end{equation}
here $\mu =\frac{m_{t}^{2}}{m_{h}^{2}}$, $\Gamma ^{2}=\frac{\Gamma
_{t}^{2}}{m_{h}^{2}}$ and $\gamma ^{2}=\frac{\Gamma
_{W}^{2}}{m_{h}^{2}}$, with $\Gamma_t\approx1.28$ GeV and
$\Gamma_W\approx2.14$ GeV are the SM full width decay for top quark
and $W$ boson, respectively \cite{pdg}. The three body decay rate is
given by the formula

\begin{equation}
d\Gamma _{h\longrightarrow
W\overline{b}c}=\frac{\overline{\left\vert \mathcal{M}\right\vert
}^{2}}{2m_{h}}\left[ \frac{d^{3}\overrightarrow{p_{1}}}{\left(
2\pi \right) ^{3}2E_{1}}\right] \left[ \frac{d^{3}\overrightarrow{p_{2}}}{%
\left( 2\pi \right) ^{3}2E_{2}}\right] \left[ \frac{d^{3}\overrightarrow{%
p_{3}}}{\left( 2\pi \right) ^{3}2E_{3}}\right] \left( 2\pi \right)
^{4}\delta ^{4}\left( p-p_{1}-p_{2}-p_{3}\right).  \label{golden}
\end{equation}
Using the delta function to perform the $\overrightarrow{p_{3}}$
integral and setting the polar axis along $\overrightarrow{p_{1}}$,
we have
\begin{equation}
\Gamma _{h\longrightarrow W\overline{b}c}=\Gamma_{11}+\Gamma_{22}+%
\Gamma_{12},  \label{g11}
\end{equation}
where we define
\begin{equation}
\Gamma _{11}=\frac{m_{h}}{256\pi ^{3}}\int
\int_{R_{xy}}\overline{\left\vert \mathcal{M}_{1}\right\vert
}^{2}dxdy, \label{g11}
\end{equation}
\begin{equation}
\Gamma _{22}=\frac{m_{h}}{256\pi ^{3}}\int
\int_{R_{xy}}\overline{\left\vert \mathcal{M}_{2}\right\vert
}^{2}dxdy, \label{g22}
\end{equation}
and
\begin{equation}
\Gamma _{12}=\frac{m_{h}}{256\pi ^{3}}\int \int_{R_{xy}}\left( \overline{%
\mathcal{M}_{1}^{\dagger}\mathcal{M}_{2}}+\overline{\mathcal{M}_{2}^{\dagger
}\mathcal{M}_{1}}\right)dxdy, \label{g12}
\end{equation}
with the $R_{xy}$\ region is defined by
\begin{equation}
\frac{1}{2}\left( 2-x-\sqrt{x^{2}-4\mu _{1}}\right) \leq y\leq \frac{1}{2}%
\left( 2-x+\sqrt{x^{2}-4\mu _{1}}\right)
\end{equation}
and
\begin{equation}
2\sqrt{\mu _{1}}\leq x\leq 1+\mu _{1}.
\end{equation}
We can obtain the next results whether we write equations
(\ref{m11f}), (\ref{m22f}), (\ref{m1m2f}) and (\ref{m2m1f}) with
dimensionless parameters and substitute in equations (\ref{g11}),
(\ref{g22}) and (\ref{g12}),
\begin{equation}
\Gamma _{11}=\frac{2g^{2}m_{h}}{\left( 16\pi \right) ^{3}}\left[
\left\vert S_{231}^{u}-P_{231}^{u}\right\vert ^{2}\left(
J_{1}+J_{2}-J_{4}-J_{6}\right)
+\left\vert S_{231}^{u}+P_{231}^{u}\right\vert ^{2}\left( J_{3}+J_{5}\right) %
\right] ,
\end{equation}%
\begin{equation}
\Gamma _{22}=\frac{g^{4}q_{11}^{2}\left\vert V_{cb}\right\vert ^{2}m_{h}}{%
512\pi ^{3}}\left[ J_{7}+J_{8}\right] ,
\end{equation}%
\begin{equation}
\Gamma _{12}=\frac{\left\vert S_{231}^{u}+P_{231}^{u}\right\vert
g^{3}q_{11}V_{cb}m_{h}}{256\pi ^{3}}\left( J_{9}\sin \theta
+J_{10}\cos \theta +J_{11}\sin \theta +J_{12}\cos \theta \right) ,
\end{equation}%
and%
\begin{equation}
\widetilde{\Gamma }_{12}=\frac{\left\vert
S_{231}^{u}+P_{231}^{u}\right\vert g^{3}q_{11}V_{cb}m_{h}}{256\pi
^{3}}\left( J_{9}\sin \theta -J_{10}\cos \theta +J_{11}\sin \theta
-J_{12}\cos \theta \right) .
\end{equation}%
The $J_{i}$ integrals, for $i=1,...,12$, are given by%

\begin{figure}[tbp]
\centering
\includegraphics[scale=0.75]{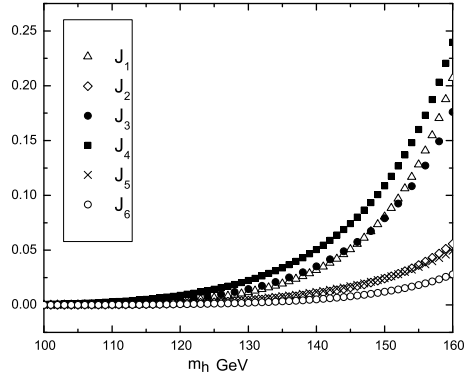}\\
\caption{Graphics for the integrals of the $\Gamma_{11}$.}
\label{j1to6}
\end{figure}

\begin{figure}[tbp]
\centering
\includegraphics[scale=0.75]{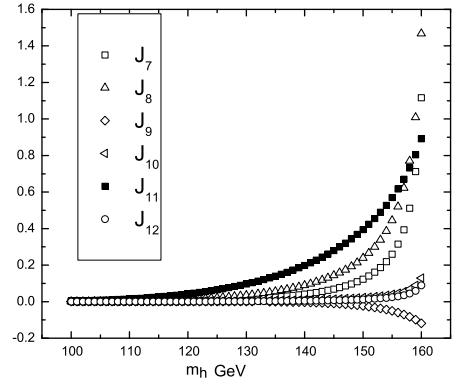}\\
\caption{Graphics for the integrals of the $\Gamma_{22}$ and
$\Gamma_{12}$.} \label{j7to12}
\end{figure}
\begin{equation}
J_{1}=\frac{1}{\mu _{1}}\int \int_{R_{xy}} \frac{%
\left( x+y-\mu _{1}-1\right) \left( x+y+\mu _{1}-1\right) \left(
2-x-y\right) }{\left( x+y-1-\mu \right) ^{2}+\mu \Gamma ^{2}}dxdy,
\label{j1}
\end{equation}
\begin{equation}
J_{2}=\int \int_{R_{xy}} \frac{\left( x+y-\mu _{1}-1\right) \left(
2-x-y\right) }{\left( x+y-1-\mu \right) ^{2}+\mu \Gamma ^{2}}dxdy
\label{j2}
\end{equation}
\begin{equation}
J_{3}=\frac{\mu }{\mu _{1}}\int \int_{R_{xy}} \frac{%
\left( x+y-\mu _{1}-1\right) \left( 1-y-\mu _{1}\right) }{\left(
x+y-1-\mu \right) ^{2}+\mu \Gamma ^{2}}dxdy,  \label{j3}
\end{equation}%
\begin{equation}
J_{4}=\frac{1}{\mu _{1}}\int \int_{R_{xy}} \frac{%
\left( x+y-1\right) \left( x+y-\mu _{1}-1\right) \left( 1-y-\mu
_{1}\right) }{\left( x+y-1-\mu \right) ^{2}+\mu \Gamma ^{2}}dxdy,
\label{j4}
\end{equation}%
\begin{equation}
J_{5}=\mu \int \int_{R_{xy}} \frac{1-x+\mu _{1}}{%
\left( x+y-1-\mu \right) ^{2}+\mu \Gamma ^{2}}dxdy,  \label{j5}
\end{equation}%
\begin{equation}
J_{6}=\int \int_{R_{xy}} \frac{\left( x+y-1\right) \left( 1-x+\mu _{1}\right) }{%
\left( x+y-1-\mu \right) ^{2}+\mu \Gamma ^{2}}dxdy,  \label{j6}
\end{equation}%
\begin{equation}
J_{7}=\mu _{1}\int \int_{R_{xy}} \frac{\left( 1-x+\mu _{1}\right)
}{\left( 1-x\right) ^{2}+\mu _{1}\gamma ^{2}}dxdy, \label{j7}
\end{equation}%
\begin{equation}
J_{8}=\int \int_{R_{xy}} \frac{\left( x+y+\mu _{1}-1\right) \left(
1-y-\mu _{1}\right) }{\left( 1-x\right) ^{2}+\mu _{1}\gamma
^{2}}dxdy, \label{j8}
\end{equation}%
\begin{equation}
J_{9}=\sqrt{\mu \mu _{1}}\int \int_{R_{xy}} \frac{%
\left( 1-x+\mu _{1}\right) \left[ \left( x+y-1-\mu \right) \left(
1-x\right) +\sqrt{\mu \mu _{1}}\gamma \Gamma \right] }{\left[ \left(
x+y-1-\mu \right)
^{2}+\mu \Gamma ^{2}\right] \left[ \left( 1-x\right) ^{2}+\mu _{1}\gamma ^{2}%
\right] }dxdy,  \label{j9}
\end{equation}%
\begin{equation}
J_{10}=\sqrt{\mu \mu _{1}}\int \int_{R_{xy}} \frac{%
\left( 1-x+\mu _{1}\right) \left[ \sqrt{\mu }\Gamma \left( 1-x\right) -\sqrt{%
\mu _{1}}\gamma \left( x+y-1-\mu \right) \right] }{\left[ \left(
x+y-1-\mu \right) ^{2}+\mu \Gamma ^{2}\right] \left[ \left(
1-x\right) ^{2}+\mu _{1}\gamma ^{2}\right] }dxdy,  \label{j10}
\end{equation}%
\begin{equation}
J_{11}=\sqrt{\frac{\mu }{\mu _{1}}}\int \int_{R_{xy}} \frac{%
\left( x+y+\mu _{1}-1\right) \left( 1-y-\mu _{1}\right) \left[
\left( x+y-1-\mu \right) \left( 1-x\right) +\sqrt{\mu \mu
_{1}}\gamma \Gamma \right] }{\left[ \left( x+y-1-\mu \right)
^{2}+\mu \Gamma ^{2}\right] \left[ \left( 1-x\right) ^{2}+\mu
_{1}\gamma ^{2}\right] }dxdy,  \label{j11}
\end{equation}%
\begin{equation}
J_{12}=\sqrt{\frac{\mu }{\mu _{1}}}\int \int_{R_{xy}} \frac{%
\left( x+y+\mu _{1}-1\right) \left( 1-y-\mu _{1}\right) \left[ \sqrt{\mu }%
\Gamma \left( 1-x\right) -\sqrt{\mu _{1}}\gamma \left( x+y-1-\mu \right) %
\right] }{\left[ \left( x+y-1-\mu \right) ^{2}+\mu \Gamma
^{2}\right] \left[ \left( 1-x\right) ^{2}+\mu _{1}\gamma ^{2}\right]
}dxdy.  \label{j12}
\end{equation}
The graphics for the $J_i$, $i=1,...,12$, are shown in the figures
\ref{j1to6} and \ref{j7to12}.


\begin{references}

\bibitem{Gunion:1989we}
  J.~F.~Gunion, H.~E.~Haber, G.~L.~Kane and S.~Dawson

\bibitem{Erler:2010wa}
  J.~Erler,
  Phys.\ Rev.\  D {\bf 81}, 051301 (2010) arXiv:1002.1320 [hep-ph].

\bibitem{Flacher:2008zq}
  H.~Flacher, M.~Goebel, J.~Haller, A.~Hocker, K.~Moenig and J.~Stelzer,
  Eur.\ Phys.\ J.\  C {\bf 60}, 543 (2009) [arXiv:0811.0009 [hep-ph]].

\bibitem{DiazCruz:2003qs}
  J.~L.~Diaz-Cruz and D.~A.~Lopez-Falcon,
  Phys.\ Lett.\  B {\bf 568}, 245 (2003)
  [arXiv:hep-ph/0304212].

\bibitem{Baur:2002gp}
U. Baur,eConfC010630:P1WG1,2001. hep-ph/0202001

\bibitem{Carena:2002es}
  M.~S.~Carena and H.~E.~Haber,
  Prog.\ Part.\ Nucl.\ Phys.\  {\bf 50}, 63 (2003)
  [arXiv:hep-ph/0208209].

\bibitem{Nath:2010zj}
  P.~Nath {\it et al.},
  Nucl.\ Phys.\ Proc.\ Suppl.\  {\bf 200-202}, 185 (2010)
  [arXiv:1001.2693 [hep-ph]].

\bibitem{Bustamante:2009us}
  M.~Bustamante, L.~Cieri and J.~Ellis,
  arXiv:0911.4409 [hep-ph].

\bibitem{Ellis:2010wx}
  J.~Ellis,
  Int.\ J.\ Mod.\ Phys.\  A {\bf 25}, 2409 (2010)
  [arXiv:1004.0648 [hep-ph]].

\bibitem{susyrev}
See, for instance, recent reviews in ``Perspectives on
Supersymmetry'', ed. G.\,L. Kane, World Scientific Publishing Co.,
1998; H.\,E. Haber, Nucl. Phys. Proc. Suppl. {\bf 101}, 217 (2001),
[hep-ph/0103095.]

\bibitem{Haber:2000jh}
H. E. Haber, Nucl. Phys. Proc. Suppl. \textbf{101}
217 (2001), hep-ph/0103095.

\bibitem{ArkaniHamed:2001nc}
N. Arkani-Hamed, A. Cohen and H. Georgi, Phys. Lett. B{\bf 513}
(2001) 232 [arXiv: hep-h/0105239].

\bibitem{Aranda:2007tg}
  A.~Aranda, J.~L.~Diaz-Cruz, J.~Hernandez-Sanchez and R.~Noriega-Papaqui,
  Phys.\ Lett.\  B {\bf 658}, 57 (2007)
  [arXiv:0708.3821 [hep-ph]].

\bibitem{Aranda:2002dz}
  A.~Aranda, C.~Balazs and J.~L.~Diaz-Cruz,
  Nucl.\ Phys.\  B {\bf 670}, 90 (2003)
  [arXiv:hep-ph/0212133].

\bibitem{Chang:2010et}
  W.~F.~Chang, J.~N.~Ng and A.~P.~Spray,
  arXiv:1004.2953 [hep-ph].

\bibitem{Iltan:2007zz}
  E.~O.~Iltan,
  Eur.\ Phys.\ J.\  C {\bf 51}, 689 (2007)
  [arXiv:hep-ph/0511241].

\bibitem{Haber:1978jt}
  H.~E.~Haber, G.~L.~Kane and T.~Sterling,
  Nucl.\ Phys.\  B {\bf 161}, 493 (1979).

\bibitem{Liu:1987ng}
  J.~Liu and L.~Wolfenstein,
  Nucl.\ Phys.\  B {\bf 289}, 1 (1987).

\bibitem{Wu:1994ja}
  Y.~L.~Wu and L.~Wolfenstein,
  Phys.\ Rev.\ Lett.\  {\bf 73}, 1762 (1994)
  [arXiv:hep-ph/9409421].

\bibitem{Carena:2000yx}
M. Carena and others, Report of the Tevatron Higgs working group
(2000), hep-ph/0010338.

\bibitem{Ginzburg:2002wt}
  I.~F.~Ginzburg, M.~Krawczyk and P.~Osland,
  arXiv:hep-ph/0211371.

 \bibitem{Ginzburg:2004vp}
  I.~F.~Ginzburg and M.~Krawczyk,
  Phys.\ Rev.\  D {\bf 72}, 115013 (2005)
  [arXiv:hep-ph/0408011].

\bibitem{Accomando:2006ga}
  E.~Accomando {\it et al.},
  arXiv:hep-ph/0608079.

\bibitem{Glashow:1976nt}
  S.~L.~Glashow and S.~Weinberg,
  Phys.\ Rev.\  D {\bf 15}, 1958 (1977).

\bibitem{DiazCruz:2004tr}
  J.~L.~Diaz-Cruz, R.~Noriega-Papaqui and A.~Rosado,
  Phys.\ Rev.\  D {\bf 69}, 095002 (2004)
  [arXiv:hep-ph/0401194].

\bibitem{DiazCruz:2007be}
  J.~L.~Diaz-Cruz,
  Phys.\ Rev.\ Lett.\  {\bf 100}, 221802 (2008)
  [arXiv:0711.0488 [hep-ph]].


\bibitem{Atwood:2005bf}
  D.~Atwood, S.~Bar-Shalom and A.~Soni,
  Phys.\ Lett.\  B {\bf 635}, 112 (2006)
  [arXiv:hep-ph/0502234].

\bibitem{DiazCruz:2004ss}
  J.~L.~Diaz-Cruz,
  Mod.\ Phys.\ Lett.\  A {\bf 20}, 2397 (2005)
  [arXiv:hep-ph/0409216].

\bibitem{Aranda:2005st}
  A.~Aranda, J.~L.~Diaz-Cruz and A.~Rosado,
  Int.\ J.\ Mod.\ Phys.\  A {\bf 22}, 1417 (2007)
  [arXiv:hep-ph/0507230].

\bibitem{DiazCruz:2006ki}
  J.~L.~Diaz-Cruz and A.~Rosado,
  Rev.\ Mex.\ Fis.\  {\bf 53}, 396 (2007)
  [arXiv:hep-ph/0610167].

\bibitem{Barbieri:2005kf}
  R.~Barbieri and L.~J.~Hall,
  arXiv:hep-ph/0510243.

\bibitem{Froggatt:2007qp}
  C.~D.~Froggatt, R.~Nevzorov, H.~B.~Nielsen and D.~Thompson,
  Phys.\ Lett.\  B {\bf 657}, 95 (2007)
  [arXiv:0708.2903 [hep-ph]].

\bibitem{Ginzburg:2005yw}
  I.~F.~Ginzburg,
  Acta Phys.\ Polon.\  B {\bf 37}, 1161 (2006)
  [arXiv:hep-ph/0512102].

\bibitem{Maniatis:2007vn}
  M.~Maniatis, A.~von Manteuffel and O.~Nachtmann,
  Eur.\ Phys.\ J.\  C {\bf 57}, 719 (2008)
  [arXiv:0707.3344 [hep-ph]].

\bibitem{Gerard:2007kn}
  J.~M.~Gerard and M.~Herquet,
  Phys.\ Rev.\ Lett.\  {\bf 98}, 251802 (2007)
  [arXiv:hep-ph/0703051].

\bibitem{ElKaffas:2007rq}
  A.~W.~El Kaffas, P.~Osland and O.~M.~Ogreid,
  Nonlin.\ Phenom.\ Complex Syst.\  {\bf 10}, 347 (2007)
  [arXiv:hep-ph/0702097].

\bibitem{Fritzsch:1977za}
H. Fritzsch, Phys. Lett. \textbf{B}70 (1977) 436.

\bibitem{Cheng:1987rs}
  T.~P.~Cheng and M.~Sher,
  Phys.\ Rev.\  D {\bf 35}, 3484 (1987).

\bibitem{Carcamo:2006dp}
  A.~E.~Carcamo Hernandez, R.~Martinez and J.~A.~Rodriguez,
  Eur.\ Phys.\ J.\  C {\bf 50}, 935 (2007)
  [arXiv:hep-ph/0606190].

\bibitem{Zhou:2003kd}
  Y.~F.~Zhou,
  J.\ Phys.\ G {\bf 30}, 783 (2004)
  [arXiv:hep-ph/0307240].

\bibitem{Aoki:2009ha}
  M.~Aoki, S.~Kanemura, K.~Tsumura and K.~Yagyu,
  Phys.\ Rev.\  D {\bf 80}, 015017 (2009)
  [arXiv:0902.4665 [hep-ph]].


\bibitem{Logan:2010ag}
  H.~E.~Logan and D.~MacLennan,
  Phys.\ Rev.\  D {\bf 81}, 075016 (2010)
  [arXiv:1002.4916 [hep-ph]].

\bibitem{Logan:2009uf}
  H.~E.~Logan and D.~MacLennan,
  Phys.\ Rev.\  D {\bf 79}, 115022 (2009)
  [arXiv:0903.2246 [hep-ph]].

\bibitem{Gunion:2005ja}
  J.~F.~Gunion and H.~E.~Haber,
  Phys.\ Rev.\  D {\bf 72}, 095002 (2005)
  [arXiv:hep-ph/0506227].

\bibitem{DiazCruz:1992uw}
  J.~L.~Diaz-Cruz and A.~Mendez,
  Nucl.\ Phys.\  B {\bf 380}, 39 (1992).

\bibitem{Lee:1973iz}
T.D. Lee, 
Phys. Rev. D8, 1226 (1973).

\bibitem{Hall:1981bc}
L. J. Hall and M. B. Wise, Nucl. Phys. \textbf{B}187, 397, (1981).

\bibitem{Donoghue:1978cj}
J.F. Donoghue and L. F. Li, 
Phys. Rev. \textbf{D}19, 945 (1979).



\bibitem{Osland:2008aw}
  P.~Osland, P.~N.~Pandita and L.~Selbuz,
  Phys.\ Rev.\  D {\bf 78}, 015003 (2008)
  [arXiv:0802.0060 [hep-ph]].

\bibitem{DiazCruz:2002er}
  J.~L.~Diaz-Cruz,
  JHEP {\bf 0305}, 036 (2003)
  [arXiv:hep-ph/0207030].

\bibitem{FN}
C. D. Frogatt and H. B. Nielsen, Nucl. Phys. B{\bf 147}, 277 (1979).

\bibitem{ourwork1}
  J.~L.~Diaz-Cruz and G.~Lopez Castro,
  Phys.\ Lett.\  B {\bf 301}, 405 (1993);

  J.~L.~Diaz-Cruz, R.~Noriega-Papaqui and A.~Rosado,
  Phys.\ Rev.\  D {\bf 71}, 015014 (2005)
  [arXiv:hep-ph/0410391];

  A.~Aranda, J.~Lorenzo Diaz-Cruz, J.~Hernandez-Sanchez and E.~Ma,
  Phys.\ Rev.\  D {\bf 81}, 075010 (2010)
  [arXiv:1001.4057 [hep-ph]];

\bibitem{ourwork2}
  J.~L.~Diaz-Cruz and J.~J.~Toscano,
  Phys.\ Rev.\  D {\bf 62}, 116005 (2000)
  [arXiv:hep-ph/9910233];

  U.~Cotti, J.~L.~Diaz-Cruz, R.~Gaitan, H.~Gonzales and A.~Hernandez-Galeana,
  Phys.\ Rev.\  D {\bf 66}, 015004 (2002)
  [arXiv:hep-ph/0205170];

  J.~L.~Diaz-Cruz, D.~K.~Ghosh and S.~Moretti,
  Phys.\ Lett.\  B {\bf 679}, 376 (2009)
  [arXiv:0809.5158 [hep-ph]];

\bibitem{ourwork3}
  J.~L.~Diaz-Cruz, J.~Hernandez--Sanchez, S.~Moretti, R.~Noriega-Papaqui and A.~Rosado,
  Phys.\ Rev.\  D {\bf 79}, 095025 (2009)
  [arXiv:0902.4490 [hep-ph]];

  C.~Balazs, J.~L.~Diaz-Cruz, H.~J.~He, T.~M.~P.~Tait and C.~P.~Yuan,
  Phys.\ Rev.\  D {\bf 59}, 055016 (1999)
  [arXiv:hep-ph/9807349];

  J.~L.~Diaz-Cruz and O.~A.~Sampayo,
  Int.\ J.\ Mod.\ Phys.\  A {\bf 8}, 4339 (1993);

  J.~L.~Diaz-Cruz and O.~A.~Sampayo,
  Phys.\ Rev.\  D {\bf 50}, 6820 (1994).


\bibitem{otherswork1}
  A.~Arhrib, M.~Capdequi Peyranere, W.~Hollik and S.~Penaranda,
  arXiv:hep-ph/0307391;

  L.~Randall,
  JHEP {\bf 0802}, 084 (2008)
  [arXiv:0711.4360 [hep-ph]];

  S.~Bejar, J.~Guasch and J.~Sola,
  Nucl.\ Phys.\  B {\bf 675}, 270 (2003)
  [arXiv:hep-ph/0307144];

  A.~Arhrib,
  Phys.\ Rev.\  D {\bf 72}, 075016 (2005)
  [arXiv:hep-ph/0510107].

\bibitem{Pich:2009sp}
A. Pich and P. Tuzon, Phys. Rev. \textbf{D}80 091702 (2009),
hep-ph/09081554.


\bibitem{Jung:2010ik},
M. Jung, A. Pich and P. Tuzon, hep-ph/10060470

\bibitem{Braeuninger:2010td}
Braeuninger, Carolin B. and Ibarra, Alejandro and Simonetto,
Cristoforo, Phys. Lett. \textbf{B}692 189 (2010), hep-ph/1005.5706

\bibitem{PortugalMinHix}
  N.~Barros e Sa, A.~Barroso, P.~Ferreira and R.~Santos,
  PoS C {\bf HARGED2008}, 014 (2008)
  [arXiv:0906.5453 [hep-ph]];

  A.~Barroso, P.~M.~Ferreira and R.~Santos,
  Phys.\ Lett.\  B {\bf 652}, 181 (2007)
  [arXiv:hep-ph/0702098];

  A.~Barroso, P.~M.~Ferreira and R.~Santos,
  Afr.\ J.\ Math.\ Phys.\  {\bf 3}, 103 (2006)
  [arXiv:hep-ph/0507329];

  A.~Barroso, P.~M.~Ferreira and R.~Santos,
  Phys.\ Lett.\  B {\bf 632}, 684 (2006)
  [arXiv:hep-ph/0507224].


\bibitem{Ivanov:2006yq}
  I.~P.~Ivanov,
  Phys.\ Rev.\  D {\bf 75}, 035001 (2007)
  [Erratum-ibid.\  D {\bf 76}, 039902 (2007)]
  [arXiv:hep-ph/0609018].


\bibitem{Maniatis:2006fs}
  M.~Maniatis, A.~von Manteuffel, O.~Nachtmann and F.~Nagel,
  Eur.\ Phys.\ J.\  C {\bf 48}, 805 (2006)
  [arXiv:hep-ph/0605184].

\bibitem{Ma:2010ya}
  E.~Ma and M.~Maniatis,
  arXiv:1005.3305 [hep-ph].

\bibitem{Haber:2006ue}
  H.~E.~Haber and D.~O'Neil,
  Phys.\ Rev.\  D {\bf 74}, 015018 (2006)
  [arXiv:hep-ph/0602242].

\bibitem{pdg}
K. Nakamura  et al. (Particle Data Group), J. Phys. G 37, 075021
(2010)

\bibitem{working-progress}
J. L. D\'iaz-Cruz, J. H. Montes de Oca Y., working in progress.

\end{references}
\end{document}